\documentclass[useAMS,usenatbib]{mnras}

\usepackage{newtxtext,newtxmath}

\usepackage[T1]{fontenc}
\usepackage{ae,aecompl}

\usepackage{graphicx}	
\usepackage{amsmath}	
\usepackage{amssymb}	
\usepackage{blkarray}
\usepackage{bigstrut}
\usepackage{layouts}
\usepackage[math]{cellspace}

\numberwithin{equation}{section}

\newcommand{\xx}{\bmath{x}}
\newcommand{\zz}{\bmath{z}}
\newcommand{\zzc}{\bmath{\bar z}}
\newcommand{\rr}{\bmath{r}}
\newcommand{\rrc}{\bmath{\bar r}}
\newcommand{\vv}{\bmath{v}}
\newcommand{\vvc}{\bmath{\bar v}}

\newcommand{\REAL}{\mathbb{R}}
\newcommand{\COMPLEX}{\mathbb{C}}

\newcommand{\AUG}[1]{\bmath{\breve{#1}}}

\newcommand{\Zz}{\AUG{z}}
\newcommand{\ZZ}{\AUG{Z}}
\newcommand{\Rr}{\AUG{r}}
\newcommand{\RR}{\AUG{R}}
\newcommand{\Vv}{\AUG{v}}
\newcommand{\VV}{\AUG{V}}
\newcommand{\augG}{\AUG{G}}

\newcommand{\vmV}{\vec{\mat{V}}}
\newcommand{\vmVH}{\vec{\mat{V}}^{*}}
\newcommand{\vmZ}{\vec{\mat{Z}}}
\newcommand{\vmZH}{\vec{\mat{Z}}^{*}}
\newcommand{\vmR}{\vec{\mat{R}}}
\newcommand{\vmRH}{\vec{\mat{R}}^{*}}

\newcommand{\vmG}{\vec{\mat{G}}}
\newcommand{\vmGH}{\vec{\mat{G}}^{*}}

\newcommand{\mat}[1]{{\mathbf{#1}}}

\newcommand{\JJ}{\mat{J}}
\newcommand{\DD}{\mat{D}}
\newcommand{\MM}{\mat{M}}
\newcommand{\GG}{\mat{G}}

\newcommand{\Matrix}[2]{\left [ \begin{array}{@{}#1@{}}#2\end{array} \right ]}

\newcommand{\Rop}[1]{\mathcal{R}_{{#1}}}
\newcommand{\Lop}[1]{\mathcal{L}_{{#1}}}

\newcommand{\VEC}[1]{\mathrm{vec}\,{#1}}

\DeclareMathOperator{\vect}{vec}

\newcommand{\bigfrac}{\dfrac{\partial{\vmV}}{\partial{\vmG}}}

\newcommand{\edits}[1]{\textcolor{black}{#1}}

\newcommand{\editss}[1]{\textcolor{black}{#1}}

\newcommand{\editsss}[1]{\textcolor{black}{#1}}

\everymath{\displaystyle}

\title[CubiCal]{CubiCal - Fast radio interferometric calibration suite exploiting complex optimisation}

\author[J. S. Kenyon et al.]{
J. S. Kenyon,$^{1}$\thanks{E-mail: jonosken@gmail.com}
O. M. Smirnov,$^{1,2}$
T. L. Grobler$^{3}$
and S. J. Perkins$^{2}$
\\
$^{1}$Department of Physics \& Electronics, Rhodes University, Grahamstown, 6140 South Africa \\
$^{2}$SKA South Africa, 3rd Floor, The Park, Park Road, Pinelands, 7405 South Africa\\
$^{3}$Department of Mathematical Sciences, Computer Science Division, Stellenbosch University, Private Bag X1, Matieland, 7602 South Africa
}

\date{Accepted 2018 May 06. Received 2018 May 03; in original form 2018 March 14}

\pubyear{2018}

\begin{document}
\label{firstpage}
\pagerange{\pageref{firstpage}--\pageref{lastpage}}
\maketitle

\begin{abstract}
It has recently been shown that radio interferometric gain calibration can be expressed succinctly in the language of complex optimisation. In addition to providing an elegant framework for further development, it exposes properties of the calibration problem which can be exploited to accelerate traditional non-linear least squares solvers such as Gauss-Newton and Levenberg-Marquardt. We extend existing derivations to chains of Jones terms: products of several gains which model different aberrant effects. In doing so, we find that the useful properties found in the single term case still hold. We also develop several specialised solvers which deal with complex gains parameterised by real values. The newly developed solvers have been implemented in a Python package called CubiCal, which uses a combination of Cython, multiprocessing and shared memory to leverage the power of modern hardware. We apply CubiCal to both simulated and real data, and perform both direction-independent and direction-dependent self-calibration. Finally, we present the results of some rudimentary profiling to show that CubiCal is competitive with respect to existing calibration tools such as MeqTrees.

\end{abstract}




\section{Introduction}


Calibration (specifically, the correction of antenna-based gain errors) of radio interferometric data is of increasing importance in the era of the Square Kilometre Array  \citep[SKA,][]{dewdney2009} and its ilk. The massive volume of data produced by these instruments \citep[e.g.][]{broekema2015} necessitates the development of faster calibration algorithms and highly optimised implementations thereof.

Using the nomenclature proposed by \cite{noordam2010}, radio interferometric calibration can be grouped into three categories: first-, second- and third-generation. First-generation calibration (1GC) makes use of calibrator field (fields containing sources with known flux, shape and spectral behaviour) observations to determine gain solutions which can be transferred to the target field. Second-generation calibration (2GC), often referred to as self-calibration or self-cal, attempts to reconcile a model of the target field with the observed visibility data by solving for the direction-independent (DI) gains. Third-generation calibration (3GC) refers to the family of techniques used to treat direction-dependent (DD) effects - individual sources may be be corrupted by effects unique to their position and these errors cannot be accurately modelled by the average gains across the field.  A more detailed description of these categories appears in \cite{smirnov2011a,smirnov2011b,smirnov2011c}.

Determining the gains requires the solution of a non-linear least squares (NLLS) problem. Traditionally, this has been accomplished by the application of techniques such as Gauss-Newton (GN) and Levenberg-Marquardt (LM) (\citealt{madsen2004} provides an overview). Both of these methods make use of gradient information to iteratively reconstruct the gains. However, prior to recent advancements in the field of complex optimisation (\citealt{kreutzdelgado2009,sorber2012}), these methods were only suitable for determining real-valued parameters. As radio interferometric gains are generally complex-valued, it has been necessary to circumvent this limitation by treating the real and imaginary parts of the gain as independent, real-valued parameters. 

Fortunately, the aforementioned advances have extended the GN and LM techniques to functions of complex variables by employing the \textit{Wirtinger} derivative \citep{wirtinger1927}. This obviates the need to split the gains into two real values and instead treats the complex-valued gains and their conjugates as independent parameters. This, in turn, exposes some useful properties in the GN and LM methods which can be exploited to accelerate the computation of the gains \citep{tasse2014, smirnov2015}. Under certain conditions, the derived update rules can be shown to be equivalent to \textit{StefCal} \citep{salvini2014}, the current state-of-the-art in fast, direction-independent calibration. The mathematics involved will be presented briefly in Section \ref{sec:maths}.

\editsss{The Wirtinger formalism, in the context of direction-dependent gain calibration, has been implemented in the {\tt killMS} package\footnote{\url{https://github.com/saopicc/killMS}} which includes both the CohJones algorithm described by \citet{smirnov2015}, and a Kalman-filter based approach \citep{tasse2014b}. This has had success with KAT-7 \citep{scaife2015}, LOFAR \citep{girard2016} and ATCA \citep{coriat2018} data, and is now a part of the LOFAR Two-Metre Sky Survey pipeline \citep{shimwell2018,tasse2018}. In this work, we develop other aspects of the mathematical framework arising from \citet{smirnov2015}.}

\editsss{Firstly, we} extend the framework  to solve for multiple gain terms. In the context of the Radio Interferometer Measurement Equation \citep[RIME,][]{hamaker1996}, this introduces the notion of a \textit{Jones Chain}: a chain of gain terms, each of which may be defined over different time and frequency solution intervals. In Section \ref{sec:chain}, we derive an expression for the parameter update with respect to any element in this chain. These chains are useful for disentangling various sources of error, as we can choose solution intervals in an attempt to capture different behaviours in time and frequency. Additionally, it is simple to combine DI terms with DD terms and express the entirety of a calibration procedure in a single chain. This strategy has a unique property; multiple gain terms may be solved sequentially without applying the inverse gains to the observed visibilities.

There are additional ways in which the methods developed here and in \cite{smirnov2015} may be taken further. In Section \ref{sec:spec} we show that by applying the chain rule, we can differentiate complex gains with respect to their parameters. This makes it possible to develop specialised solvers, the simplest example of which is phase-only. By differentiating the complex gains with respect to a real phase, we develop true phase-only calibration. In doing so, we discover a useful property of phase-only gain calibration which massively reduces its computational cost. Whilst any number of specialised solvers may be devised, the current work will be restricted to the derivation of phase-only, delay/rate and pointing error solvers.  

Section \ref{sec:impl} details the implementation of these new solvers in a Python package called CubiCal. CubiCal has been accelerated using Cython \citep{dalcin2010} which allows for the inclusion of C-like, compiled code in Python. Naturally, this code is more optimal than generic Python and provides a substantial performance boost, especially when coupled with multiprocessing and shared memory. The package also implements the \textit{AllJones} strategy presented in \cite{smirnov2015}, which not only accelerates 3GC but also makes peeling \citep{noordam2004} unnecessary, as all directions are calibrated simultaneously. The package currently implements $2\times2$ complex, phase-only and delay/rate solvers. However, CubiCal is highly modular, and additional solvers will be added in future. The model prediction component of CubiCal is handled by Montblanc \citep{perkins2015}, another Python package which implements a GPU-accelerated version of the RIME. 

CubiCal has a measurement set interface and has been applied successfully to both simulated and real data, as described in Section \ref{sec:results}. A brief discussion of CubiCal's future appears alongside our conclusions in Section \ref{sec:conc}.

\section{Mathematical Foundations}
\label{sec:maths}

This section is based on \cite{smirnov2015}, as it shares much of its mathematical background. However, for the sake of flow and notational completeness, some of the basic concepts will be reintroduced here and we will elaborate on the points more crucial to this work. 

\subsection{NLLS for complex scalar variables}
\label{ssec:scalar}

Both GN and LM make extensive use of first derivative information \citep{madsen2004}. Radio interferometric gains are, generally, complex valued. The result is that the function we wish to minimise is a function of $n$ complex variables ($f(\zz),$ $\zz\in\COMPLEX^n$). Taking the derivative with respect to these variables is not possible using conventional differentiation as the partial derivative $\partial\bar z/\partial z$ is not defined. This is usually avoided by reformulating the problem as a function of $2n$ real variables; the real and imaginary parts of the complex variable $\zz=\bmath{x}+i\bmath{y}$ are treated as independent, yielding \edits{$f(\bmath{x},\bmath{y}),$ $\bmath{x}\in\REAL^{n},\bmath{y} \in\REAL^{n}$}. 

The Wirtinger derivatives \citep{wirtinger1927},
\begin{equation}
\dfrac{\partial}{\partial z} = \dfrac{1}{2}\left ( \dfrac{\partial}{\partial x} - i\dfrac{\partial}{\partial y} \right),~~
\dfrac{\partial}{\partial \bar{z}} = \dfrac{1}{2}\left ( \dfrac{\partial}{\partial x} + i\dfrac{\partial}{\partial y} \right),
\end{equation}
treat the complex variable ($\zz$) and its conjugate ($\zzc$) as independent. The \edits{works} of \cite{kreutzdelgado2009} and \cite{sorber2012} \edits{make} use of the Wirtinger derivatives to derive complex counterparts for traditional NLLS methods, specifically GN and LM. Using these, it is possible to solve problems of the form:
\begin{equation}
\label{eq:LSmin}
\min_{\bmath{z}} ||\bmath{d}-\bmath{v}(\zz,\zzc)||_F.
\end{equation}

This is of particular interest when solving for antenna gains as it describes a least-squares fit of the data ($\bmath{d} \in \COMPLEX^m$) using a model ($\bmath{v} \in \COMPLEX^m$) parameterised by complex variables. Note that $||\cdot||_F$ is the Frobenius norm.

In order to express the problem of gain calibration in the framework of complex NLLS it is first necessary to establish some notation. If the independent variables with respect to which we wish to minimise are given by $\zz$ and $\zzc$, we define our length $2n$ {\em augmented parameter vector} as:
\begin{equation}
\label{eq:saugz}
\Zz = 
\begin{bmatrix}
\zz \\ \zzc
\end{bmatrix}.
\end{equation}

Similarly, if we define our residuals as a function of $\Zz$ as $\rr(\Zz)$ and $\rrc(\Zz)$, we define our length $2m$ {\em augmented residual vector} as:
\begin{equation}
\Rr = 
\begin{bmatrix}
\rr(\Zz) \\ \rrc(\Zz)
\end{bmatrix}.
\end{equation}

In the context of radio interferometry, it is common to define the Jacobian (matrix of first derivatives) with respect to the model $\vv(\Zz)$ and its conjugate $\vvc(\Zz)$. Consequently we define the length $2m$ (consistent with the dimension of $\Rr$) {\em augmented model vector} as:
\begin{equation}
\label{eq:saugv}
\Vv = 
\begin{bmatrix}
\vv(\Zz) \\ \vvc(\Zz)
\end{bmatrix}.
\end{equation}

Armed with the definitions above, it is possible express the Jacobian, $\JJ$ as:
\begin{equation}
\label{eq:Jkrk}
\renewcommand{\arraystretch}{1.5}
\JJ = \dfrac{\partial \Vv}{\partial \Zz} = 
\cellspacetoplimit 3pt
\cellspacebottomlimit 3pt
\setlength{\arraycolsep}{4pt}
\begin{bmatrix}
\dfrac{\partial{\vv}}{\partial{\zz}} & \dfrac{\partial{\vv}}{\partial{\zzc}} \\ \dfrac{\partial{\vvc}}{\partial{\zz}} & \dfrac{\partial{\vvc}}{\partial{\zzc}}
\end{bmatrix}
= 
\Matrix{cc}{\JJ_{\vv\zz} & \JJ_{\vv\zzc} \\ \JJ_{\vvc\zz} & \JJ_{\vvc\zzc} },
\end{equation}
where, $\JJ_{\vv\zz}$, $\JJ_{\vv\zzc}$, $\JJ_{\vvc\zz}$ and $\JJ_{\vvc\zzc}$ are referred to as the partial Jacobians of the problem. Each partial Jacobian has dimensions $m \times n$, and the complete Jacobian has dimensions $2m \times 2n$. Note that the diagonally adjacent partial Jacobians are element-wise conjugates of each other.  

The above is sufficient to express the GN and LM update rules in terms of the complex Jacobian. As these are iterative methods, at each iteration we find the update, $\delta \Zz$, to the augmented parameter vector. The update rules for GN and LM \edits{\citep{madsen2004}} are given by:
\begin{equation}
\label{eq:GN}
\delta\Zz = (\JJ^H \JJ)^{-1}\JJ^H \Rr, 
\end{equation}
and
\begin{equation}
\label{eq:LM}
\delta\Zz = (\JJ^H \JJ + \lambda\DD)^{-1}\JJ^H \Rr,
\end{equation}
respectively, where $\DD$ is a diagonal matrix containing the diagonal entries of $\JJ^H \JJ$ and $\lambda$ is a damping parameter which steers the update between GN ($\lambda = 0$) and steepest descent ($\lambda = \infty$). Due to the dependence of both $\JJ$ and $\Rr$ on $\Zz$, every term in the update rules will vary with iteration. The iteration index is omitted for the sake of simplicity.

\subsection{NLLS for complex $2\times2$ variables}
\label{ssec:polar}

Thus far, we have only presented the mathematics pertaining to the scalar problem and have not considered the case of $2\times2$ complex variables. This is of particular importance in the field of radio interferometry as observations often include polarisation information. In order to deal with this case, \cite{smirnov2015} presented an operator calculus to derive NLLS updates with respect to $2\times2$ complex variables. 

A matrix operator can be defined as a function, $\mathcal{F}$, which maps one complex-valued $2\times2$ matrix to another:
\begin{equation}
\mathcal{F}:\COMPLEX^{2\times2} \to \COMPLEX^{2\times2}.
\end{equation}

An operator can be applied to a matrix $\mat{Z}$ to produce a new matrix $\mat{Y}$ or, more concisely:
\begin{equation}
\mat{Y}=\mathcal{F}\hspace{0.15em}\mat{Z}=\mathcal{F}\hspace{0.15em}[\mat{Z}].
\end{equation}

Two such operators will be particularly useful - the so-called left and right multipliers. Their function is relatively simple; given $2\times2$ matrices $\mat{A}$ and $\mat{B}$, the left and right multipliers are:
\begin{equation}
\begin{array}{l@{~}l}
\Lop{\mat{A}}\mat{B} &= \mat{AB} \\
\Rop{\mat{A}}\mat{B} &= \mat{BA}
\end{array}.
\end{equation}

A detailed explanation of these operators and their derivation appears in appendix B of \cite{smirnov2015}. Here, we will merely state some useful properties of these operators and note that there exists an isomorphism, $\mathbb{W}$, which allows us to express our $2\times2$ operators as $4\times4$ matrices acting on vectors:
\begin{equation}
\label{eq:vecrels}
\begin{array}{l@{~}l@{~}l}
\VEC{(\Lop{\mat{A}}\mat{B})} &\equiv \VEC{(\mat{AB})} &\equiv \mathbb{W}\Lop{\mat{A}}\VEC{(\mat{B})} \\
\VEC{(\Rop{\mat{A}}\mat{B})} &\equiv \VEC{(\mat{BA})} &\equiv \mathbb{W}\Rop{\mat{A}}\VEC{(\mat{B})}
\end{array},
\end{equation}
where:
\begin{equation}
\label{eq:iso}
\begin{array}{l@{~}l}
\mathbb{W}\Lop{\mat{A}} &= \mat{I} \otimes \mat{A} \\
\mathbb{W}\Rop{\mat{A}} &= \mat{A}^T \otimes \mat{I}
\end{array}~~.
\end{equation}

The useful properties of these operators are quite intuitive and can be inferred from their definitions. The first such property is the manner in which similar operators can be combined:

\begin{equation}
\Lop{\mat{A}}\Lop{\mat{B}} = \Lop{\mat{AB}},~~~\Rop{\mat{A}}\Rop{\mat{B}} = \Rop{\mat{BA}}.
\end{equation}

The second is that the inverse of the operator is equivalent to applying the inverse of its argument:
\begin{equation}
(\Lop{\mat{A}})^{-1} = \Lop{\mat{A^{-1}}},~~~(\Rop{\mat{A}})^{-1} = \Rop{\mat{A^{-1}}}.
\end{equation}

We are now in an excellent position to reformulate the contents of Section \ref{ssec:scalar} with respect to $2\times2$ matrices. In order to do so, we introduce the notion of a vector of matrices. Whilst this may seem peculiar at first, it is quite simple; where conventionally the entries of a vector would be scalar, we have $2\times2$ matrices instead. Given $N$ potentially complex-valued $2\times2$ matrices, $( \mat{Z}_1,\dots,\mat{Z}_{\scriptstyle N})$, we adopt the following notation:

\begin{equation}
\label{eq:2x2vec}
\vec{\mat{Z}} = 
\begin{bmatrix}
\mat{Z}_1,\dots,\mat{Z}_{\scriptstyle N}
\end{bmatrix}^T,
\end{equation}
where $\vec{\mat{Z}}$ denotes a vector of matrices. We define its element-wise conjugate transpose as follows:

\begin{equation}
\label{eq:2x2vech}
\vmZH = 
\begin{bmatrix}
\mat{Z}^H_1,\dots,\mat{Z}^H_{\scriptstyle N}
\end{bmatrix}^T.
\end{equation}

We are now free to write down the $2\times2$ equivalents of equations \ref{eq:saugz} through \ref{eq:saugv}:

\begin{equation}
\label{eq:paugz}
\ZZ = 
\begin{bmatrix} 
\vmZ \\ \vmZH 
\end{bmatrix},
~~~\RR = 
\begin{bmatrix} 
\vmR(\ZZ) \\ \vmRH(\ZZ) 
\end{bmatrix},
~~~\VV = 
\begin{bmatrix}
\vmV(\ZZ) \\ \vmVH(\ZZ)
\end{bmatrix}.
\end{equation}

The Jacobian can be written in terms of these augmented vectors of matrices:

\begin{equation}
\label{eq:pJkrk}
\JJ = \dfrac{\partial \VV}{\partial \ZZ} = 
\begin{bmatrix}
\dfrac{\partial{\vmV}}{\partial{\vmZ}} & \dfrac{\partial{\vmV}}{\partial{\vmZH}} \\ \dfrac{\partial{\vmVH}}{\partial{\vmZ}} & \dfrac{\partial{\vmVH}}{\partial{\vmZH}}
\end{bmatrix}
= 
\begin{bmatrix}
\JJ_{\vmV\vmZ} & \JJ_{\vmV\vmZH} \\ 
\JJ_{\vmVH\vmZ} & \JJ_{\vmVH\vmZH}
\end{bmatrix}.
\end{equation}

The derivatives which appear in equation \ref{eq:pJkrk} require some explanation. For a length $m$ vector of $2\times2$ matrices, $\vmV$, and a length $n$ vector of $2\times2$ matrices, $\vmZ$, the derivative of $\vmV$ with respect to $\vmZ$ can be interpreted as each $2\times2$ element in $\vmV$ being differentiated with respect to each $2\times2$ element in $\vmZ$. Of course, matrix-by-matrix differentiation is not very meaningful in general. This problem can be circumvented by vectorising (reducing to one dimension) both $\vmV$ and $\vmZ$ and performing vector-by-vector differentiation. However, as shown in appendix B1 of \cite{smirnov2015}, the operator calculus we have introduced provides an alternative approach. 

For $2\times2$ matrices $\mat{A}$, $\mat{B}$ and $\mat{C}$ we have the following:

\begin{equation}
\label{eq:pderiv}
\dfrac{\partial (\mat{ABC})}{\partial \mat{A}} = \Rop{\mat{BC}},~~~\dfrac{\partial (\mat{ABC})}{\partial \mat{B}} = \Lop{\mat{A}}\Rop{\mat{C}},~~~\dfrac{\partial (\mat{ABC})}{\partial \mat{C}} = \Lop{\mat{AB}}.
\end{equation}

These derivatives are sufficient to express the entries of the Jacobian in terms of the $2\times2$ left and right multiply operators and we can write down the $2\times2$ GN and LM equivalents:

\begin{equation}
\label{eq:pGN}
\delta\ZZ = (\JJ^H \JJ)^{-1}\JJ^H \RR, 
\end{equation}
and
\begin{equation}
\label{eq:pLM}
\delta\ZZ = (\JJ^H \JJ + \lambda\DD)^{-1}\JJ^H \RR.
\end{equation}

\section{Calibrating a Jones Chain}
\label{sec:chain}

The mathematics of of the preceding section, in particular that of Section \ref{ssec:polar}, can be employed to derive calibration with respect to specific terms in a \textit{Jones chain} (as introduced by \cite{smirnov2011a}).

\subsection{Direction-independent chains}
An interferometric measurement has contributions from two antennas (those forming the corresponding baseline). Each antenna has an associated set of Jones terms, indexed by $p$ and $q$, where $p,q \in[1,\ldots,N_{A}]$, and $N_A$ is the total number of antennas in the array. These Jones terms can be used to model various effects on the signal.

The visibility observed by a baseline $pq$ can be expressed in terms of the $2 \times 2$, full-polarisation RIME with multiple Jones terms:
\begin{equation}
\label{eq:pRIME}
\DD_{pq} = \bigg(\prod_{n=1}^{N_{J}}\GG_p^{(n)}\bigg) \MM_{pq} \bigg(\prod_{n=1}^{N_{J}}\GG_q^{(n)}\bigg)^H + \mat{N}_{pq}.
\end{equation}

In equation \ref{eq:pRIME}, $\DD_{pq}$ is the $2\times2$ visibility matrix observed by baseline $pq$, $\MM_{pq}$ is the predicted (model) visibility associated with baseline $pq$ and $\mat{N}_{pq}$ is a noise term. The $N_J$ direction independent Jones terms associated with antenna $p$ are represented by the product of $\GG_p$ terms. We also make the following useful property explicit:

\begin{equation}
\bigg(\prod_{n=1}^{N_{J}}\GG_q^{(n)}\bigg)^H = ~\bigg(\prod_{n=N_J}^{1}\GG_q^{(n)H}\bigg).
\end{equation}

We can now express the full-polarisation NLLS problem as:
\begin{equation}
\label{eq:cal:DI:pol}
\min_{\{\GG_p^{(n)}\}}\sum_{pq}||\mat{R}_{pq}||_F,~~
\mat{R}_{pq} = \DD_{pq}-\mat{V}_{pq},
\end{equation}
where,
\begin{equation}
\label{eq:pmodel}
\mat{V}_{pq} = \bigg(\prod_{n=1}^{N_{J}}\GG_p^{(n)}\bigg) \MM_{pq} \bigg(\prod_{n=1}^{N_{J}}\GG_q^{(n)}\bigg)^H.
\end{equation}

It is important to emphasise that we will not be minimising with respect to all the Jones terms simultaneously. Instead, we will minimise with respect to a single term in the chain. \edits{This amounts to solving for a subset of the parameters. However, NLLS only assures local convergence, and there is no guarantee that solving for a single chain element at a time will yield the same local minima as solving the overall problem. Therefore, the results should be treated carefully in practice. Of course, the global problem itself is also subject to local minima and the same precautions apply.} 

The update rules we derive can be applied to any term and in any order. As a result, we establish a couple of calibration strategies which we will touch on at the end of this section.

\subsection{NLLS building blocks}

Making use of equation \ref{eq:pmodel}, we can begin to derive the components of our NLLS update. First and foremost is the need to construct the elements of the Jacobian. For a single baseline, we can write the derivative of $\mat{V}_{pq}$ with respect to the $j$'th Jones term and its conjugate as:

\begin{equation}
\label{eq:pvgderiv}
\dfrac{\partial \mat{V}_{pq}}{\partial \mat{G}_p^{(j)}} = \Lop{(\prod_{n<j}\mat{G}^{(n)}_p)} \Rop{(\prod_{n>j}\mat{G}^{(n)}_p) \MM_{pq} (\prod_{n=1}^{N_J}\mat{G}^{(n)}_q)^H},
\end{equation}
and,
\begin{equation}
\label{eq:pvghderiv}
\dfrac{\partial \mat{V}_{pq}}{\partial \mat{G}_q^{(j)H}} = \Lop{(\prod_{n=1}^{N_J}\mat{G}^{(n)}_p) \MM_{pq} (\prod_{n>j}\mat{G}^{(n)}_q)^H} \Rop{(\prod_{n<j}\mat{G}^{(n)}_q)^H}. 
\end{equation}

\editss{These expressions (and many of those to follow) may seem complicated at first, but they are substantially less cumbersome than reverting to $4\times4$ expressions via equations \ref{eq:vecrels} and \ref{eq:iso}. However, it should be stressed that both representations are equally valid and, ultimately, equivalent.}

Using equations \ref{eq:pJkrk}, \ref{eq:pvgderiv} and \ref{eq:pvghderiv}, we can write out an analytic expression for the Jacobian:

\begin{equation}
\label{eq:pJacobian}
\JJ = \dfrac{\partial \VV}{\partial \augG} =
\begin{bmatrix}
\dfrac{\partial{\vmV}}{\partial{\vmG}} & \dfrac{\partial{\vmV}}{\partial{\vmGH}} \\ 
\dfrac{\partial{\vmVH}}{\partial{\vmG}} & \dfrac{\partial{\vmVH}}{\partial{\vmGH}}
\end{bmatrix}
\begin{matrix}
\scriptstyle \bigg\}~[pq]=1,\dots,N_\mathrm{bl} ~ (p<q)
\vphantom{\bigfrac} \\
\scriptstyle \bigg\}~[pq]=1,\dots,N_\mathrm{bl} ~ (p<q)
\vphantom{\bigfrac}
\end{matrix}~~,
\end{equation}
and its constituent partial Jacobians, the rows and columns of which are indexed by $a$ and $b$:
\begin{equation}
\label{eq:pjterms}
\begin{split}
\dfrac{\partial{\vmV}}{\partial{\vmG}} & = \begin{bmatrix}\Lop{(\prod_{n<j}\mat{G}^{(n)}_p)} \Rop{(\prod_{n>j}\mat{G}^{(n)}_p) \MM_{pq} (\prod_{n=1}^{N_J}\mat{G}^{(n)}_q)^H} \delta_p^b \end{bmatrix} \\
\dfrac{\partial{\vmV}}{\partial{\vmGH}} & = \begin{bmatrix}\Lop{(\prod_{n=1}^{N_J}\mat{G}^{(n)}_p) \MM_{pq} (\prod_{n>j}\mat{G}^{(n)}_q)^H} \Rop{(\prod_{n<j}\mat{G}^{(n)}_q)^H} \delta_q^b\end{bmatrix} \\
\dfrac{\partial{\vmVH}}{\partial{\vmG}} & = \begin{bmatrix}\Lop{(\prod_{n<j}\mat{G}^{(n)}_q)} \Rop{(\prod_{n>j}\mat{G}^{(n)}_q) \MM_{pq}^H (\prod_{n=1}^{N_J}\mat{G}^{(n)}_p)^H} \delta_q^b\end{bmatrix} \\
\dfrac{\partial{\vmVH}}{\partial{\vmGH}} & = \begin{bmatrix}\Lop{(\prod_{n=1}^{N_J}\mat{G}^{(n)}_q) \MM_{pq}^H (\prod_{n>j}\mat{G}^{(n)}_p)^H} \Rop{(\prod_{n<j}\mat{G}^{(n)}_p)^H} \delta_p^b\end{bmatrix}
\end{split}.
\end{equation}

Each partial Jacobian in equation \ref{eq:pJacobian} is a block with $N_\mathrm{bl}$ (number of baselines) rows and $N_\mathrm{A}$ (number of antennas) columns. \edits{Each row in these blocks is associated with a specific baseline. These baselines are identified by $pq$}; a compound index which enumerates all combinations of $p$ and $q$ such that $p<q$. $\delta_p^b$ has its usual meaning as the Dirac delta ($\delta = 1$ if $p = b$ else $\delta = 0$) \edits{and captures the fact that for each row (baseline), only certain columns will have non-zero entries}. Each entry of a partial Jacobian consists of $2\times2$ operators.

By modifying the compound $pq$ index (as was done in \cite{smirnov2015}) to include all combinations of $pq$ such that $p \neq q$ (all baselines and their conjugates), it is possible to express the Jacobian in a simpler form:

\begin{equation}
\label{eq:psimjacobian}
\JJ =
\begin{bmatrix}
\dfrac{\partial{\vmV}}{\partial{\vmG}} & \dfrac{\partial{\vmV}}{\partial{\vmGH}}
\end{bmatrix}
\begin{matrix}  
\scriptstyle \bigg\}~[pq]=1,\dots,2N_\mathrm{bl} ~ (p \neq q)
\end{matrix}~~.
\end{equation}

The components retain their definitions from equation \ref{eq:pjterms} - we are merely exploiting the fact that the vertically adjacent partial Jacobians produce the same set of expressions when $pq$ includes both baselines ($p < q$) and their conjugates ($q < p$). This can be verified by substitution.

The above observation applies equally to other components of the problem and we are free to redefine our augmented residual vector in terms of the new compound baseline index:

\begin{equation}
\label{eq:simaugr}
\RR =
\begin{bmatrix}
\mat{R}_{pq}
\end{bmatrix}
\begin{matrix}
\scriptstyle \big\}~[pq]=1,\dots,2N_\mathrm{bl} ~ (p \neq q)
\end{matrix}~~.
\end{equation}

In this instance, $\RR$ is still a vector of matrices but it is constructed by stacking the residual elements for all baselines and their conjugates into a vector of $2\times2$ matrices.

\subsection{Deriving an update rule}
\label{jc:updaterule}

Expressions \ref{eq:psimjacobian} and \ref{eq:simaugr} are the basic building blocks we needed to construct the NLLS updates. All that remains is to combine them. To do so, we need an expression for the conjugate transpose of the Jacobian. This is given by:

\begin{equation}
\label{eq:psimjacobianh}
\JJ^{H} =
\begin{bmatrix}
\bigg(\dfrac{\partial{\vmV}}{\partial{\vmG}} \bigg)^H \\ \bigg( \dfrac{\partial{\vmV}}{\partial{\vmGH}} \bigg)^H
\end{bmatrix}
\end{equation}
where, noting again that the rows and columns of the partial Jacobians are indexed by $a$ and $b$, we have:
\begin{equation}
\label{eq:pjhterms}
\begin{split}
\bigg(\dfrac{\partial{\vmV}}{\partial{\vmG}} \bigg)^H  & = \begin{bmatrix}\Lop{(\prod_{n<j}\mat{G}^{(n)}_p)^H} \Rop{(\prod_{n=1}^{N_J}\mat{G}^{(n)}_q) \MM_{pq}^H (\prod_{n>j}\mat{G}^{(n)}_p)^H} \delta_p^a\end{bmatrix} \\
\bigg( \dfrac{\partial{\vmV}}{\partial{\vmGH}} \bigg)^H & = \begin{bmatrix}\Lop{(\prod_{n>j}\mat{G}^{(n)}_q) \MM_{pq}^H (\prod_{n=1}^{N_J}\mat{G}^{(n)}_p)^H} \Rop{(\prod_{n<j}\mat{G}^{(n)}_q)} \delta_q^a\end{bmatrix}
\end{split}.
\end{equation}

We also construct the $\JJ^H\JJ$ term, which is an approximation of the Hessian, $\mat{H}$. However, due to the length of the expressions involved, it is practical to first define the following substitution:
\begin{equation}
\label{eq:suby}
\mat{Y}_{pq} = \bigg(\prod_{n>j}\mat{G}^{(n)}_p\bigg) \MM_{pq} \bigg(\prod_{n=1}^{N_J}\mat{G}^{(n)}_q\bigg)^H.
\end{equation} 

Using this substitution, we can write the following:
\begin{equation}
\label{eq:pjhj}
\mat{H} = 
\JJ^H\JJ =
\begin{bmatrix}
\mat{A} & \mat{B} \\
\mat{C} & \mat{D}
\end{bmatrix},
\end{equation}
where:
\begin{equation}
\label{eq:pjhjterms}
\begin{split}
\mat{A} & = 
\begin{cases} 
\displaystyle     \Lop{(\prod_{n<j}\mat{G}^{(n)}_a)^H(\prod_{n<j}\mat{G}^{(n)}_a)} \sum_{q \neq a}\Rop{\mat{Y}_{aq}\mat{Y}_{aq}^H} & a = b \\
      0 & a \neq b 
\end{cases}
\\
\mat{B} & = \begin{cases} 
     \Lop{(\prod_{n<j}\mat{G}^{(n)}_a)^H\mat{Y}_{ba}^H} \Rop{(\prod_{n<j}\mat{G}^{(n)}_b)^H \mat{Y}_{ab}^H} & a \neq b \\
      0 & a = b 
\end{cases}
\\
\mat{C} & = \begin{cases} 
\Rop{\mat{Y}_{ba} (\prod_{n<j}\mat{G}^{(n)}_a)} \Lop{\mat{Y}_{ab} (\prod_{n<j}\mat{G}^{(n)}_b)} & a \neq b \\
      0 & a = b 
\end{cases}
\\
\mat{D} & = 
\begin{cases} 
\displaystyle     \Rop{(\prod_{n<j}\mat{G}^{(n)}_a)^H(\prod_{n<j}\mat{G}^{(n)}_a)} 
\sum_{p \neq a} \Lop{\mat{Y}_{ap} \mat{Y}_{ap}^H} & a = b \\
      0 & a \neq b 
\end{cases}
\end{split}.
\end{equation}

Finally, we apply $\JJ^H$ to $\RR$ to obtain:

\begin{equation}
\label{eq:pjhr}
\JJ^H\RR =
\begin{bmatrix}
\mat{E} \\
\mat{F}
\end{bmatrix},
\end{equation}
where,
\begin{equation}
\begin{split}
\mat{E} & = \sum_{q \neq a} \bigg(\prod_{n<j}\mat{G}^{(n)}_a\bigg)^H \mat{R}_{aq} \mat{Y}_{aq}^H \\
\mat{F} & = \sum_{p \neq a} \mat{Y}_{ap} \mat{R}_{pa} \bigg(\prod_{n<j}\mat{G}^{(n)}_a\bigg)
\end{split}.
\end{equation}

It should be apparent that the top and bottom halves of $\JJ^H\RR$ are Hermitian with respect to each other. 

Equations \ref{eq:pjhj} and \ref{eq:pjhr} are the components of the GN and LM methods expressed in terms of Jones chains. Before going further, we note some properties of the parameter update and the approximations we make.

The first observation is that, as in the non-chain case, we need not compute the entirety of the update vector. Instead, as half the equations are the conjugate transpose of the others, we are free to reduce our parameter update to:
\begin{equation}
\label{eq:valt}
\delta\vmG = (\JJ^H \JJ)^{-1}_\mathrm{U}\JJ^H \RR, 
\end{equation}
where $(\cdot)_\mathrm{U}$ denotes the upper half a of a matrix.

Following on from this, we can exploit a property mentioned in \cite{tasse2014}: 
\begin{equation}
\label{eq:jtrick}
\VV = \JJ_{\mathrm{L}} \vmG = \frac{1}{2} \JJ \AUG{G},
\end{equation}
\edits{where $(\cdot)_\mathrm{L}$ denotes the left half of a matrix,} to show that by substituting for $\RR$ in equation \ref{eq:valt}, we obtain:
\begin{equation}
\label{eq:subr}
\delta\vmG = (\JJ^H \JJ)^{-1}_\mathrm{U} \JJ^H (\AUG{D} - \JJ_{\mathrm{L}} \vmG) = (\JJ^H \JJ)^{-1}_\mathrm{U} \JJ^H \AUG{D} - \vmG. 
\end{equation}

Noting that $\vmG_k + \delta \vmG = \vmG_{k+1}$
for iteration index $k$, we can now write the update in terms of the observed data rather than the residual:
\begin{equation}
\label{eq:halfupdatedata}
\vmG_{k+1} = (\JJ^H \JJ)^{-1}_\mathrm{U} \JJ^H \AUG{D}.
\end{equation}

Naturally, avoiding computing the residual is computationally advantageous. This property was remarked upon in \cite{smirnov2015} and holds provided
\begin{equation}
\label{eq:diagapprox}
\tilde{\mat{H}}^{-1}_\mathrm{U} \mat{H}_\mathrm{L} = \mathbb{I}~,
\end{equation}
where $\mat{H} = \JJ^H \JJ$. This is always true when $\tilde{\mat{H}}$ is an exact inversion of $\mat{H}$. However, if $\tilde{\mat{H}}$ is an approximation, this condition is not necessarily satisfied and using the residuals may be unavoidable.
 
In the direction independent case, we approximate $\mat{H}$ as a  diagonal matrix with $2\times2$ entries. Returning to equation \ref{eq:pjhj}, the diagonal approximation leaves us with:
\begin{equation}
\label{eq:pdiagjhj}
\tilde{\mat{H}} = 
\begin{bmatrix}
\mat{A} & \mat{0} \\
\mat{0} & \mat{D}
\end{bmatrix},
\end{equation}
where $\mat{A}$ and $\mat{D}$ are unchanged. This still satisfies the requirement of \ref{eq:diagapprox} and we are free to use the data directly in addition to the easily inverted diagonal matrix. 

No approximation is without repercussions. In this instance, the price we pay is one of iteration-to-iteration accuracy; we expect convergence to take more iterations. However, each iteration is far easier to compute - we have replaced the traditionally $\mathcal{O}(n^3)$ inversion of a matrix with an $\mathcal{O}(n)$ invertible approximation.

All of these useful properties can be combined to produce a single update rule. We can incorporate equation \ref{eq:pdiagjhj} into equation \ref{eq:halfupdatedata} by noting that the upper half of $\tilde{\mat{H}}$ contains a block of zeros. This block effectively excludes any contribution from the lower half of $\JJ^H \AUG{R}$ (see equation \ref{eq:pjhr}). As equation \ref{eq:pdiagjhj} already excluded the lower half of $\mat{H}$, we find that only the upper left quadrant (denoted by $(\cdot)_{\mathrm{UL}}$) of $\tilde{\mat{H}}$ is required. Consequently, we write the total effective update as:
\begin{equation}
\label{eq:approxupdate}
\vmG_{k+1} = \tilde{\mat{H}}^{-1}_\mathrm{UL} \JJ^H_\mathrm{L} \AUG{D}.
\end{equation}

Gathering the relevant terms from equations \ref{eq:pjhj} and \ref{eq:pjhr} and noting that we are free to replace $\AUG{R}$ with $\AUG{D}$, substitution into \ref{eq:approxupdate} gives:
\begin{equation}
\vmG_{k+1} = \mat{A}^{-1} \mat{E}.
\end{equation}

This in turn can be used to write out the per-antenna update rule\editss{, either by direct application of the operators or by reverting to the $4\times4$ expressions (and a vectorised update) using equations \ref{eq:vecrels} and \ref{eq:iso}}. Noting that some terms are eliminated due to the application of the inverse, the \editss{$2\times2$,} per-antenna update for the $j$'th term is:
\begin{equation}
\label{eq:pupdate}
\mat{G}_{a,k+1}^{(j)} = 
\bigg(\prod_{n<j}\mat{G}^{(n)}_a\bigg)^{-1}
\bigg(\sum_{q \neq a} \mat{D}_{aq} \mat{Y}_{aq}^H\bigg)
\bigg(\sum_{q \neq a} \mat{Y}_{aq} \mat{Y}_{aq}^H\bigg)^{-1}
\end{equation}

Equation \ref{eq:pupdate} can be used to compute the update with respect to any Jones term within the chain and is one of the most important contributions of this work. An additional feature of interest is that this expression allows for the solution of different Jones terms without needing to apply existing solutions to the data.

In the single Jones term case, equation \ref{eq:pupdate} reduces to:
\begin{equation}
\label{eq:pcomparison}
\mat{G}_{a,k+1} = 
\bigg(\sum_{q \neq a} \mat{D}_{aq} \mat{Y}_{aq}^H\bigg)
\bigg(\sum_{q \neq a} \mat{Y}_{aq} \mat{Y}_{aq}^H\bigg)^{-1}
\end{equation}
with:
\begin{equation}
\label{eq:simpleysub}
\mat{Y}_{aq} = \MM_{aq} \mat{G}_q^H.
\end{equation}

This expression should agree with equation 5.26 of \cite{smirnov2015}. However, due to an error in that paper stemming from an incorrectly applied operator, the order of the terms differ. Regardless, as mentioned in that paper, for a single Jones term this is equivalent to polarised StefCal \citep{salvini2014}.

\subsection{Solution intervals}
\label{ssec:jcsolint}

It is common practice in calibration to solve for gains at a lower time and/or frequency resolution than the data. This is usually done in an attempt to improve the signal-to-noise, thus improving the solution, or to prevent over-fitting. 

In terms of our calibration problem, this manifests as multiple time and/or frequency samples being incorporated into the solution of a single gain:
\begin{equation}
\min_{\{\GG_p^{(n)}\}}\sum_{pqs}||\mat{R}_{pqs}||_F,~~
\mat{R}_{pqs} = \DD_{pqs}-\mat{V}_{pqs},
\end{equation}
where,
\begin{equation}
\mat{V}_{pq} = \bigg(\prod_{n=1}^{N_{J}}\GG_p^{(n)}\bigg) \MM_{pqs} \bigg(\prod_{n=1}^{N_{J}}\GG_q^{(n)}\bigg)^H.
\end{equation}

The additional index, $s$, corresponds to the time and frequency samples and can be included in our compound baseline index. This can be propagated through the derivation to arrive at the following per antenna update rule:
\begin{equation}
\mat{G}_{a,k+1}^{(j)} = 
\bigg(\prod_{n<j}\mat{G}^{(n)}_a\bigg)^{-1}
\bigg(\sum_{q \neq a,s} \mat{D}_{aqs} \mat{Y}_{aqs}^H\bigg)
\bigg(\sum_{q \neq a,s} \mat{Y}_{aqs} \mat{Y}_{aqs}^H\bigg)^{-1},
\end{equation}
where:
\begin{equation}
\mat{Y}_{aqs} = \bigg(\prod_{n>j}\mat{G}^{(n)}_a\bigg) \MM_{aqs} \bigg(\prod_{n=1}^{N_J}\mat{G}^{(n)}_q\bigg)^H.
\end{equation}

In the case of a Jones chain, each term in the chain may require or prefer a different solution interval. Fortunately, this is completely tractable in terms of the given update rule. Term dependent solution intervals merely introduce an additional bit of housekeeping - we just ensure that the gains can be correctly broadcast against each other. 

\subsection{Weights}
\label{ssec:jcweights}

Incorporating weights into the polarised Jones chain is not a simple task in the general case as they introduce a Hadamard product which is difficult to handle using the $2\times2$ operator formalism. 

To circumvent this problem, we make a simplifying assumption; our weights can be represented by scalar matrices. The fact that these weights are commutative mean that we are free to solve the weighted least squares problem:
\begin{equation}
\min_{\{\GG_p^{(n)}\}}\sum_{pqs}||\mat{W}_{pqs}\mat{R}_{pqs}||_F,~~
\mat{R}_{pqs} = \DD_{pqs}-\mat{V}_{pqs},
\end{equation}
by simply pre-multiplying our scalar weights into both the data term ($\mat{D}$) and the predicted visibility component of the model term ($\mat{M}$, which appears in $\mat{V}$). Thereafter, the update rule for weighted least squares is completely identical to the unweighted case.

\subsection{Direction-dependence}
\label{ssec:jcdd}

Direction-dependent calibration is of increasing importance in light of the new and improved interferometers which \edits{have recently come} online. Performing direction-dependent calibration requires solving for $N_D$ (number of directions) sets of $N_A$ gains using $N_D$ sets of model visibilities (tantamount to having $N_D$ different skies). Our NLLS problem becomes:
\begin{equation}
\min_{\{\GG_{d,p}^{(n)}\}}\sum_{pqs}||\mat{R}_{pqs}||_F,~~
\mat{R}_{pqs} = \DD_{pqs}-\sum_{d=1}^{N_D}\mat{V}_{d,pqs},
\end{equation}
where:
\begin{equation}
\mat{V}_{d,pqs} = \bigg(\prod_{n=1}^{N_{J}}\GG_{d,p}^{(n)}\bigg) \MM_{d,pqs} \bigg(\prod_{n=1}^{N_{J}}\GG_{d,q}^{(n)}\bigg)^H.
\end{equation}

Traditionally, direction-dependent calibration has been performed using peeling \citep{noordam2004}. Peeling solves for each set of gains in series, subtracting off all sources not in the current direction of interest. By iterating between directions, this method converges to a gain solution per direction. This often gives excellent results, but lacks parallelism and may be impractical for large measurement sets. It is also known to produce artefacts called ghosts \citep{grobler2014} as a result of calibrating using a largely incomplete sky model per direction.

\cite{smirnov2015} incorporate direction dependent calibration into polarised gain calibration and suggest several different approximations for $\mat{H}$. In this paper we will focus on only one of those strategies - \textit{AllJones}. \textit{AllJones} is the most approximate of the presented strategies as it assumes that the problem is separable by both direction and antenna. The result is that \edits{it} has the largest potential speed-up.

This speed up is again obtained by assuming that $\mat{H}$ is diagonal, thus making it easy to invert. Intuitively, we only expect separability by direction (as enforced by the diagonal approximation) to hold when our directions are sufficiently separated on the sky. Fortunately, this seems realistic - we are unlikely to apply independent gains to sources which are close together.

There are, however, repercussions. In the case of \textit{AllJones}, the requirement of equation \ref{eq:diagapprox} is not met and we are forced to revert to using the residuals. This is unfortunate, but the huge reduction in the computational cost of inverting $\mat{H}$ is still appealing. 

\textit{CohJones}, one of the other strategies presented in \cite{smirnov2015}, only assumes separability by antenna. It fulfils the requirements of equation \ref{eq:diagapprox}, but still requires a block-wise inversion of $\mat{H}$. We elected to pursue only \textit{AllJones}, as it has several subtle advantages over \textit{CohJones} when it comes to implementation.

Returning to the NLLS update, in the \textit{AllJones} case we can write down the per-direction, per-antenna parameter update as:
\begin{equation}
\label{eq:ddpupdate}
\begin{split}
& \mat{G}_{d,a,k+1}^{(j)} = \\ 
& ~ \bigg(\prod_{n<j}\mat{G}^{(n)}_{d,a}\bigg)^{-1}
\bigg(\sum_{q \neq a,s} \mat{R}_{d,aqs} \mat{Y}_{d,aqs}^H\bigg)
\bigg(\sum_{q \neq a,s} \mat{Y}_{d,aqs} \mat{Y}_{d,aqs}^H\bigg)^{-1},
\end{split}
\end{equation}
where:
\begin{equation}
\mat{Y}_{d,aqs} = \bigg(\prod_{n>j}\mat{G}^{(n)}_{d,a}\bigg) \MM_{d,aqs} \bigg(\prod_{n=1}^{N_J}\mat{G}^{(n)}_{d,q}\bigg)^H.
\end{equation}

The use of $\mat{G}$ may seem confusing at first as it is conventional to make use of $\mat{E}$ for direction-dependent terms. However, here we are trying to maintain generality and $\mat{G}$ should be thought of as a generic complex gain which may or may not be direction-dependent.

On that note, the update rule in equation \ref{eq:ddpupdate} implies that all the Jones terms are direction-dependent. This does not have to be true. We are free to combine direction-independent and direction-dependent terms, provided the direction-dependent terms are centre-most. If we solve for a direction-independent term in a mixed chain, it is necessary to sum over directions after applying all the direction-dependent terms. Similarly, if the term of interest is direction-dependent, the direction-independent gains must be broadcast into all directions.

\subsection{Calibration strategies}

Due to the fact that our update rules are defined with respect to a single term in the chain, we have a couple of options when we want to calibrate for several terms. 

The first option is to let each term iterate to convergence, generally (but not necessarily) starting from the outermost Jones term. Practically, this is often the best option as each successive solver will have substantially improved calibration. It is also possible to augment this strategy by doing more than one pass through the terms. Subsequent passes should slightly improve the final estimates.

The second approach is to alternate rapidly between terms, performing only a couple of iterations per term before moving on. Intuitively, this seems as though it would encourage each term to capture only the gain errors which it is intended to. However, we have found that there are some practical problems with this approach as it can cause issues with convergence.

\section{Specialised Solvers}
\label{sec:spec}

The preceding section dealt with the calibration of a generic Jones chain made up of relatively arbitrary complex gains. Here, we will do precisely the opposite and present several solvers which have been specialised for different types of gain calibration.

\subsection{The chain rule}

Thus far, we have only been interested in solving the general NLLS problem with respect to complex variables. It is, however, possible to parameterise our complex variables or gains. Parameterisation can reduce the degrees of freedom, may yield more useful physical information about the system and/or force the gains to behave in specific ways. 

The first step in constructing these specialised solvers is to establish the behaviour of the Jacobian when our complex variables are a function of real-valued parameters. We modify our original statement of a general scalar NLLS problem to reflect this parametrisation:
\begin{equation}
\min_{\bmath{x}} ||\bmath{d}-\bmath{v}(\zz(\xx),\zzc(\xx))||_F,
\end{equation}
where $\xx$ is a vector of real-valued parameters on which the value of the complex variables $\zz$ and $\zzc$ depend. Our augmented model vector is consequently given by: 
\begin{equation}
\Vv = 
\begin{bmatrix}
\vv(\Zz(\xx)) \\
\vvc(\Zz(\xx))
\end{bmatrix}.
\end{equation}

The overall Jacobian in this case will be made up of the derivatives of the model term with respect to the real-valued parameters. However, we can define the overall Jacobian using the chain rule by noting that:
\begin{equation}
\JJ = \JJ_1 \JJ_2, ~~~\JJ_1 = \frac{\partial \Vv}{\partial \Zz}, ~~~\JJ_2 = \frac{\partial \Zz}{\partial \xx}.
\end{equation}

We will make use of this property to combine our existing derivation of the Jacobian with interesting parametrisations. 

\subsection{A simple phase-only solver}
\label{ssec:po}

The first and simplest specialised solver deals with gains which have a variable phase but unity amplitude. We will perform the derivation for the full-polarisation case but will require the gains to be diagonal. This is common practice when performing phase-only calibration. The derivation will also be limited to a single Jones term - there are some practical problems with using parameterised terms within a chain in the $2\times2$ formalism. A more general $4\times4$ formalism will be the topic of a future work.


Our starting point is the direction-independent, single Jones term NLLS problem:
\begin{equation}
\min_{\{\bmath{\phi}_p\}}\sum_{pq}||\mat{R}_{pq}||_F,~~
\mat{R}_{pq} = \DD_{pq}-\mat{V}_{pq},
\end{equation}
where,
\begin{equation}
\mat{V}_{pq} = \GG_p \MM_{pq} \GG_q^H,
\end{equation}
and,
\begin{equation}
\GG_p = 
\begin{bmatrix}
g_p^{XX} & 0 \\
0 & g_p^{YY}
\end{bmatrix}
= 
\begin{bmatrix}
e^{i\phi_p^{XX}} & 0 \\
0 & e^{i\phi_p^{YY}}
\end{bmatrix}.
\end{equation}

The phases are given by $\phi$ and the correlations with which they are associated are given by $XX$ and $YY$. Specifically, these refer to the on-diagonal correlations and could equivalently be $LL$ and $RR$ for circular feeds. 

In order to construct our Jacobian, we will need to compute $\JJ_1$ (derivative of $\VV$ with respect to $\augG$) and $\JJ_2$ (derivative of $\augG$ with respect to phase).

$\JJ_1$ has already been computed in equation \ref{eq:psimjacobian}. However, as that was for the Jones chain case, we restate it in its simpler, single term form here:
\begin{equation}
\label{eq:poj1}
\JJ_1 = 
\begin{bmatrix}
\mathbb{W}\Rop{\MM_{pq} \mat{G}_q^H} \delta_p^b & \mathbb{W}\Lop{\mat{G}_p \MM_{pq}}  \delta_q^b  
\end{bmatrix}
\begin{matrix}
\scriptstyle \big\}~[pq]=1,\dots,2N_\mathrm{bl} ~ (p \neq q)
\end{matrix}.
\end{equation}

The appearance of $\mathbb{W}$ in equation \ref{eq:poj1} is a necessary evil. In truth, it is not possible to take the product of two Jacobians when they are defined in terms of $2\times2$ operators. The reason is intuitive: in order to differentiate a $2\times2$ gain, it is necessary to first vectorise into a $4\times1$ vector and then differentiate with respect to the entries of the parameter vector.

We define our parameter (phase) vector as:
\begin{equation}
\vec{\Phi} = 
\begin{bmatrix}
\vec{\phi}_1, \dots, \vec{\phi}_{N_A}
\end{bmatrix}^T,
\end{equation}
where,
\begin{equation}
\vec{\phi}_p = 
\begin{bmatrix} 
\phi_p^{XX}, \phi_p^{YY}
\end{bmatrix}^T.
\end{equation}
Using this notation and adding that $N_{\mathrm{PPA}}$ is the number of parameters per antenna, we can express our second Jacobian (with dimensions $2N_{\mathrm{A}}$ by $N_{\mathrm{A}}N_{\mathrm{PPA}}$) as:
\begin{equation}
\label{eq:poj2}
\JJ_2 = \frac{\partial \augG}{\partial \vec{\Phi}} = 
\begin{bmatrix}
\frac{\partial\vec{g}_p}{\partial \vec{\phi}_b} \delta_p^b \\
\frac{\partial\vec{g}^{\edits{\dagger}}_q}{\partial \vec{\phi}_b} \delta_q^b
\end{bmatrix}
\begin{matrix}
\scriptstyle \Big\}~p=1,\dots,N_\mathrm{A} \displaystyle\vphantom{\frac{\partial\vec{g}^H_p}{\partial \vec{\phi_b}}} \\
\scriptstyle \Big\}~q=1,\dots,N_\mathrm{A} \displaystyle\vphantom{\frac{\partial\vec{g}^H_p}{\partial \vec{\phi_b}}}
\end{matrix}~~,
\end{equation}
where it is understood that:
\begin{equation}
\vec{g}_p = \vect(\mat{G}_p), ~~~ \vec{g}^{\edits{\dagger}}_p = \vect(\mat{G}^H_p).
\end{equation}

The entries of $\JJ_2$ are $4\times2$ blocks and are thus compatible with the $4\times4$ blocks of $\JJ_1$. Thus we finally take the product of the two Jacobians and obtain:
\begin{equation}
\label{eq:poj}
\JJ =
\begin{bmatrix}
\mathbb{W}\Rop{\MM_{pq} \mat{G}_q^H} \frac{\partial\vec{g}_p}{\partial \vec{\phi}_b} \delta_p^b + \mathbb{W}\Lop{\mat{G}_p \MM_{pq}} \frac{\partial\vec{g}^{\edits{\dagger}}_q}{\partial \vec{\phi}_b} \delta_q^b 
\end{bmatrix}.
\end{equation}

By combining the two Jacobians, we actually end up with a smaller overall Jacobian ($2N_\mathrm{bl} \times N_\mathrm{A}N_\mathrm{PPA}$) and the resulting matrix no longer has an obvious left and right half.

Given the simplicity of the analytic expression of $\JJ$ it is easy to write down a similar expression for $\JJ^H$:
\begin{equation}
\label{eq:pojh}
\JJ^H =
\begin{bmatrix}
\bigg(\frac{\partial\vec{g}_p}{\partial \vec{\phi}_a}\bigg)^H \mathbb{W}\Rop{\mat{G}_q \MM^H_{pq}}  \delta_p^a + \bigg(\frac{\partial\vec{g}^{\edits{\dagger}}_q}{\partial \vec{\phi}_a}\bigg)^H \mathbb{W}\Lop{\MM_{pq}^H \mat{G}_p^H}  \delta_q^a
\end{bmatrix}.
\end{equation}

The next step is to determine analytic expressions for $\JJ^H\JJ$ and $\JJ^H \RR$. Due to the change in the dimensions of $\JJ$, $\JJ^H\JJ$ will have shape $N_\mathrm{A}N_\mathrm{PPA} \times N_\mathrm{A}N_\mathrm{PPA}$ and is no longer comprised of four distinct blocks:
\begin{equation}
\label{eq:pojhj}
\mat{H} = 
\JJ^H\JJ =
\begin{bmatrix}
\mat{A}
\end{bmatrix},
\end{equation}
where,
\begin{equation}
\label{eq:pojhjterms}
\begin{split}
\mat{A} = 
\begin{cases}
\displaystyle
\sum_{q \neq a}\bigg(\frac{\partial\vec{g}_a}{\partial \vec{\phi}_a}\bigg)^H \mathbb{W}\Rop{\MM_{aq} \mat{G}_q^H \mat{G}_q \MM^H_{aq}} \frac{\partial\vec{g}_a}{\partial \vec{\phi}_a} ~~ + \\[1em]
\displaystyle
\sum_{p \neq a} \bigg(\frac{\partial\vec{g}^{\edits{\dagger}}_a}{\partial \vec{\phi}_a}\bigg)^H \mathbb{W}\Lop{\MM_{pa}^H \mat{G}_p^H \mat{G}_p \MM_{pa}} \frac{\partial\vec{g}^{\edits{\dagger}}_a}{\partial \vec{\phi}_a}  & a = b \\[1em]
\bigg(\frac{\partial\vec{g}_a}{\partial \vec{\phi}_a}\bigg)^H \mathbb{W}\Rop{\mat{G}_b \MM^H_{ab}} \mathbb{W}\Lop{\mat{G}_a \MM_{ab}} \frac{\partial\vec{g}^{\edits{\dagger}}_b}{\partial \vec{\phi}_b} ~~ + \\[1em]
\bigg(\frac{\partial\vec{g}^{\edits{\dagger}}_a}{\partial \vec{\phi}_a}\bigg)^H \mathbb{W}\Lop{\MM_{ba}^H \mat{G}_b^H} \mathbb{W}\Rop{\MM_{ba} \mat{G}_a^H} \frac{\partial\vec{g}_b}{\partial \vec{\phi}_b}
& a \neq b 
\end{cases}
\end{split}.
\end{equation}

The entries of $\mat{H}$ are in fact $2 \times 2$ matrices. This somewhat surprising return to $2 \times 2$ blocks is a result of multiplication with the second Jacobian, the entries of which are $4 \times 2$ matrices.

$\JJ^H \RR$ will have dimension $N_\mathrm{A}N_\mathrm{PPA} \times 1$ with each entry being a $2 \times 1$ matrix - this is consistent with the $2 \times 2$ blocks of $\mat{H}$:
\begin{equation}
\begin{split}
\JJ^H\RR =
\displaystyle
&\left[\sum_{q \neq a}\bigg(\frac{\partial\vec{g}_a}{\partial \vec{\phi_a}}\bigg)^H \mathbb{W}\Rop{\mat{G}_q \MM^H_{aq}} \mat{R}_{aq} ~~ + \right. \\ & \qquad \left. \sum_{p \neq a}\bigg(\frac{\partial\vec{g}^{\edits{\dagger}}_a}{\partial \vec{\phi_a}}\bigg)^H \mathbb{W}\Lop{\MM_{pa}^H \mat{G}_p^H} \mat{R}_{pa}\right]
\end{split}~~.
\end{equation}
This is equivalent to:
\begin{equation}
\label{eq:povecjhr}
\begin{split}
\JJ^H\RR =
\displaystyle
&\left[\sum_{q \neq a}\bigg(\frac{\partial\vec{g}_a}{\partial \vec{\phi_a}}\bigg)^H \vect\big(\mat{R}_{aq} \mat{G}_q \MM^H_{aq}\big) ~~ + \right. \\ &\qquad \left. \sum_{p \neq a}\bigg(\frac{\partial\vec{g}^{\edits{\dagger}}_a}{\partial \vec{\phi_a}}\bigg)^H \vect\big(\MM_{pa}^H \mat{G}_p^H \mat{R}_{pa}\big)\right]
\end{split}~~.
\end{equation}

In order to simplify this expression further, we need some insight into the structure of the phase-derivative terms. Those appearing in equation \ref{eq:povecjhr} are given by:

\begin{equation}
\label{eq:poderivs}
\begin{split}
\bigg(\frac{\partial\vec{g}_a}{\partial \vec{\phi_a}}\bigg)^H &= 
\begin{bmatrix}
-i\overline{g}_a^{XX}&0&0&0 \\
0&0&0&-i\overline{g}_a^{YY}
\end{bmatrix} \\
\bigg(\frac{\partial\vec{g}_a^{\edits{\dagger}}}{\partial \vec{\phi_a}}\bigg)^H &=
\begin{bmatrix}
\phantom{-}ig^{XX}_a&0&0&0 \\
0&0&0&\phantom{-}ig^{YY}_a
\end{bmatrix}
\end{split}~~.
\end{equation}

From these expressions, it is clear that only the first and last elements of the vectorised component will be included in the result. We now make a small mathematical leap to write out another equivalent expression:
\begin{equation}
\label{eq:podiagjhr}
\begin{split}
\JJ^H\RR =
\displaystyle
&\left[\sum_{q \neq a} \mathrm{diag}(-i\mat{G}_a^H \odot (\mat{R}_{aq} \mat{G}_q \MM^H_{aq})) ~~+ \right. \\ &\qquad \left. \sum_{p \neq a} \mathrm{diag}(i\mat{G}_a \odot (\MM_{pa}^H \mat{G}_p^H \mat{R}_{pa}))\right]
\end{split}~~,
\end{equation}
where $\mathrm{diag}(\cdot)$ extracts a vector containing the diagonal entries of its argument. Equation \ref{eq:podiagjhr} also makes it possible to see \edits{that the} second summation is the conjugate of the first. Exploiting the fact that $z + \overline{z} = 2\operatorname{\mathbb{R}e}(z)$, we can write:
\begin{equation}
\JJ^H\RR =
\begin{bmatrix}
\displaystyle
\sum_{q \neq a} \mathrm{diag}(2\operatorname{\mathbb{R}e}(-i\mat{G}_a^H \odot (\mat{R}_{aq} \mat{G}_q \MM^H_{aq})))
\end{bmatrix}.
\end{equation}

Finally, noting that $\operatorname{\mathbb{R}e}(-iz) = \operatorname{\mathbb{I}m}(z)$, we arrive at a simple expression for $\JJ^H\RR$ in terms of $2\times2$ matrices:
\begin{equation}
\label{eq:pojhrfinal}
\JJ^H\RR =
\begin{bmatrix}
\displaystyle
\sum_{q \neq a} \mathrm{diag}(2\operatorname{\mathbb{I}m}(\mat{G}_a^H \odot (\mat{R}_{aq} \mat{G}_q \MM^H_{aq})))
\end{bmatrix}.
\end{equation}

Equations \ref{eq:pojhj} and \ref{eq:pojhrfinal} are sufficient to make use of the GN and LM methods. However, once again we first make some approximations to improve the performance of the algorithms. 

The first step is to diagonalise $\mat{H}$. This is simply the component of equation \ref{eq:pojhjterms} for which $a=b$. This expression is still unnecessarily complicated and can be greatly simplified by noting that for diagonal, phase-only gains:
\begin{equation}
\mat{G}_p^H \mat{G}_p = \mat{G}_p \mat{G}_p^H = \mat{I}.
\end{equation}
Using this, in conjunction with the phase derivatives presented in equation \ref{eq:poderivs}, the approximation of $\mat{H}$ becomes:

\begin{equation}
\mat{\tilde{H}} = 
\begin{bmatrix}
\displaystyle
2\sum_{q \neq a} \mat{I} \odot (\MM_{aq} \MM^H_{aq}) 
\end{bmatrix}.
\end{equation}

The fact that $\tilde{\mat{H}}$ is invariant with respect to the gains is a remarkable result. As a direct consequence, $\tilde{\mat{H}}$ need only be computed (and inverted) once, removing it from the iterative component of the solution procedure. This, in turn, speeds up  individual iterations substantially. 

Whilst diagonalisation makes a dramatic difference in this case, it is possible to go further. The trick of equation \ref{eq:jtrick} doesn't hold in this case but it is possible to obtain a similar result. By substituting $\mat{D}_{pq} - \mat{G}_p\mat{M}_{pq}\mat{G}_q^H$ for $\mat{R}_{pq}$ into the expression for $\JJ^H\RR$, it is possible to show that:
\begin{equation}
\JJ^H\RR =
\begin{bmatrix}
\displaystyle
\sum_{q \neq a} \mathrm{diag}(2\operatorname{\mathbb{I}m}(\mat{G}_a^H \odot (\mat{D}_{aq} \mat{G}_q \MM^H_{aq})))
\end{bmatrix},
\end{equation}
as all terms not including the data, $\mat{D}_{pq}$, are zero. Thus, we are once again free to use the data directly, obviating the costly residual computation.

Combining these approximations, we can write down the per antenna, phase-only update rule as:
\begin{equation}
\partial\vec{\phi}_a = 
\bigg(\sum_{q \neq a} \mat{I} \odot (\MM_{aq} \MM^H_{aq})\bigg)^{-1}
\bigg(\sum_{q \neq a} \mathrm{diag}(\operatorname{\mathbb{I}m}(\mat{G}_a^H \odot \mat{D}_{aq} \mat{G}_q \MM^H_{aq})) \bigg),
\end{equation}
noting that we compute the change in the phases, not in the gains. We update the gains to be consistent with the updated phases. 

The extensions to the update rule made in Sections \ref{ssec:jcsolint}, \ref{ssec:jcweights} and \ref{ssec:jcdd} can also be applied to the phase-only case. We omit the additional expressions here for the sake of brevity.

\subsection{A note on parameterised gains}
\label{ssec:paramgains}

Having derived the update rule for the phase-only case, i.e. each antenna gain parameterised by a single phase, it is interesting to see if a different approach to the same problem can produce similar results. 

The aim is to infer the update rule for a parameterised gain given what we already know about the case for a general complex gain. To do so, we return to the basic GN update rule in the case where $\JJ = \JJ_1\JJ_2$ for an arbitrary parameter vector $\vec{x}$:
\begin{equation}
\label{eq:param:gn}
\partial{\vec{x}} = (\JJ^H\JJ)^{-1}\JJ^H\RR = (\JJ_2^H\JJ_1^H\JJ_1\JJ_2)^{-1}\JJ_2^H\JJ_1^H\RR. 
\end{equation}

We can now group the terms in equation \ref{eq:param:gn}:
\begin{equation}
\label{eq:param:gngrouped}
\partial{\vec{x}} = (\JJ_2^H(\JJ_1^H\JJ_1)\JJ_2)^{-1}\JJ_2^H(\JJ_1^H\RR) = (\JJ_2^H\tilde{\mat{H}}_1\JJ_2)^{-1}\JJ_2^H(\JJ_1^H\RR), 
\end{equation}
and replace $\JJ_1^H\JJ_1$ with its diagonal approximation which we derived in Section \ref{jc:updaterule}, noting that we will need to cast it into its $4\times4$ matrix form. 
For a single Jones term, this is given by the following $4\times4$ block diagonal matrix:
\begin{equation}
\label{eq:param:happrox}
\tilde{\mat{H}}  =
\begin{bmatrix}
\displaystyle \sum_{q \neq a}  \mathbb{W}\Rop{\MM_{aq} \mat{G}_q^H \mat{G}_q \MM_{aq}^H} & \mat{0} \\
\mat{0} & \displaystyle \sum_{p \neq a}\mathbb{W}\Lop{\MM_{pa}^H \mat{G}_p^H \mat{G}_p \MM_{pa}}
\end{bmatrix}.
\end{equation}

For an arbitrary parameter vector, $\JJ_2$ is given by:
\begin{equation}
\JJ_2 =
\begin{bmatrix}
\frac{\partial\vec{g}_p}{\partial \vec{x}_b} \\
\frac{\partial\vec{g}^{\edits{\dagger}}_q}{\partial \vec{x}_b}
\end{bmatrix}
\begin{matrix}
\scriptstyle \Big\}~p=1,\dots,N_\mathrm{A}
\displaystyle\vphantom{\frac{\partial\vec{g}^H_q}{\partial \vec{x}_b}}\\
\scriptstyle \Big\}~q=1,\dots,N_\mathrm{A}
\displaystyle\vphantom{\frac{\partial\vec{g}^H_q}{\partial \vec{x}_b}}
\end{matrix}~~,
\end{equation}
where each element is a $4\times N_{\mathrm{PPA}}$ block and the parameter vector associated with antenna $n$ is:
\begin{equation}
\vec{x}_n = 
\begin{bmatrix}
x_n^{1}, x_n^{2}, \dots, x_n^{N_\mathrm{PPA}}
\end{bmatrix}^T.
\end{equation}

Multiplying $\JJ_2^H$ and $\JJ_2$ into $\tilde{\mat{H}}$ gives:
\begin{equation}
\label{eq:param:totaljhj}
\begin{split}
\tilde{\mat{H}}  =
\displaystyle
&\left[\sum_{q \neq a} \bigg(\frac{\partial\vec{g}^{\phantom{H}}_a}{\partial \vec{x}_a}\bigg)^H \mathbb{W}\Rop{\MM_{aq} \mat{G}_q^H \mat{G}_q \MM_{aq}^H} \frac{\partial\vec{g}_a}{\partial \vec{x}_a} \right. ~~+ \\ & \left. \qquad \displaystyle \sum_{p \neq a} \bigg(\frac{\partial\vec{g}^{\edits{\dagger}}_a}{\partial \vec{x}_a}\bigg)^H
 \mathbb{W}\Lop{\MM_{pa}^H \mat{G}_p^H \mat{G}_p \MM_{pa}}
\frac{\partial\vec{g}^{\edits{\dagger}}_a}{\partial \vec{x}_a}
\right]
\end{split}~~.
\end{equation}

This expression is completely equivalent to the diagonal component of equation \ref{eq:pojhjterms} if we let $\vec{x}_a = \vec{\phi}_a$. This is a general result - for any parameterisation of a complex gain, assuming a diagonal approximation of $\JJ_1$, the resulting approximation of the Hessian, $\tilde{\mat{H}}$, will be a diagonal $N_{\mathrm{A}} \times N_{\mathrm{A}}$ matrix comprised of $N_{\mathrm{PPA}} \times N_{\mathrm{PPA}}$ blocks. Thus, all parametrisations will yield an easily invertible block diagonal matrix.

We can also generate $\JJ^H\RR$ using a similar strategy as was used for $\tilde{\mat{H}}$. Returning to equation \ref{eq:pjhr} and simplifying it down to the single term case (in vectorised form), we obtain:
\begin{equation}
\label{eq:param:chainjhr}
\JJ^H_1\RR =
\begin{bmatrix}
\sum_{q \neq a} \vect( \mat{R}_{aq} \mat{G}_q \MM_{aq}^H) \\
\sum_{p \neq a} \vect( \MM_{pa}^H \mat{G}_p^H \mat{R}_{pa})
\end{bmatrix}.
\end{equation}

Multiplication by $\JJ^H_2$ then yields:
\begin{equation}
\label{eq:param:overalljhr}
\begin{split}
\JJ^H\RR = \JJ^H_2\JJ^H_1\RR =
&\left[\sum_{q \neq a}\bigg(\frac{\partial\vec{g}_a}{\partial \vec{x}_a}\bigg)^H \vect(\mat{R}_{aq} \mat{G}_q \MM^H_{aq}) ~~+ \right. \\ &\qquad \left. \sum_{p \neq a}\bigg(\frac{\partial\vec{g}^{\edits{\dagger}}_a}{\partial \vec{x}_a}\bigg)^H \vect(\MM_{pa}^H \mat{G}_p^H \mat{R}_{pa}) \right]
\end{split}~~.
\end{equation}

Again, this expression is equivalent to \ref{eq:povecjhr} in the case where $\vec{x}_a = \vec{\phi}_a$. In fact, we can go one step further. Equation \ref{eq:param:overalljhr} shows us the the two terms which make up $\JJ^H\RR$ are always conjugates of each other. Consequently we can use the same property as before to write:
\begin{equation}
\label{eq:param:overalljhrsimplified}
\JJ^H\RR =
\begin{bmatrix}
\sum_{q \neq a} 2\operatorname{\mathbb{R}e}\bigg(\bigg(\frac{\partial\vec{g}_a}{\partial \vec{x}_a}\bigg)^H \vect(\mat{R}_{aq} \mat{G}_q \MM^H_{aq})\bigg)
\end{bmatrix}.
\end{equation}

Provided we can write an analytic expression for the second Jacobian, we can employ this short-cut for any parameterisation. 

\subsection{Solving for phase-slopes}
\label{ssec:slopes}
A true phase-only solver is a useful tool, but it is possible to parameterise the phase in more interesting and powerful ways. For example, delay errors induce a phase slope across frequency. Consequently, delay calibration consists of solving for only two parameters - offset and slope - over a large number of channels. Both delay and rate (phase slopes in time) calibration are commonly used in the field of VLBI (Very Long Baseline Interferometry) \citep{cotton1995}. We can use the results and insights of Section \ref{ssec:paramgains} to derive a solver for this case.

For generality, we consider the case of independent slopes across time and frequency with a single intercept. Update rules for the simpler cases of either a time or frequency slope are merely simplifications of this more general approach. \edits{In order to simplify the notation, we perform the derivation for a single time and frequency. This will be amended in the final result.} We express the NLLS problem as:
\begin{equation}
\min_{\{\bmath{s}_p\}}\sum_{pq}||\mat{R}_{pq}||_F,~~
\mat{R}_{pq} = \DD_{pq}-\mat{V}_{pq},
\end{equation}
where,
\begin{equation}
\mat{V}_{pq} = \GG_p \MM_{pq} \GG_q^H,
\end{equation}
and,
\begin{equation}
\begin{split}
\GG_p &= 
\begin{bmatrix}
g_p^{XX} & 0 \\
0 & g_p^{YY}
\end{bmatrix} \\
&= 
\begin{bmatrix}
e^{i(a_p^{XX}\nu + b_p^{XX}t + c_p^{XX})} & 0 \\
0 & e^{i(a_p^{YY}\nu  + b_p^{YY}t + c_p^{YY})}
\end{bmatrix}.
\end{split}
\end{equation}

The parameter vector contains the coefficients which describe the per-antenna phase slopes and is given by:
\begin{equation}
\vec{{\edits{\Gamma}}} = 
\begin{bmatrix}
\vec{{\edits{\gamma}}}_1, \dots, \vec{{\edits{\gamma}}}_{N_\mathrm{A}}
\end{bmatrix}^T 
= 
\begin{bmatrix}
\vec{a}_1, \vec{b}_1, \vec{c}_1, \dots, \vec{a}_{N_\mathrm{A}}, \vec{b}_{N_\mathrm{A}}, \vec{c}_{N_\mathrm{A}}
\end{bmatrix}^T,
\end{equation}
where,
\begin{equation}
\vec{a}_p = 
\begin{bmatrix}
a_p^{XX}, a_p^{YY}
\end{bmatrix}^T, 
~~~ 
\vec{b}_p = 
\begin{bmatrix}
b_p^{XX}, b_p^{YY}\end{bmatrix}^T, 
~~~ 
\vec{c}_p = 
\begin{bmatrix}
c_p^{XX}, c_p^{YY}
\end{bmatrix}^T.
\end{equation}

At this juncture, we return to the analytic expressions for $\tilde{\mat{H}}$ and $\JJ^H\RR$ (equations \ref{eq:param:totaljhj} and \ref{eq:param:overalljhr}) where we substitute for the elements of the parameter vector to obtain:
\begin{equation}
\label{eq:slopes:totaljhj}
\begin{split}
\tilde{\mat{H}}  =
&\left[\sum_{q \neq a} \bigg(\frac{\partial\vec{g}^{\phantom{H}}_a}{\partial \vec{{\edits{\gamma}}}_a}\bigg)^H \mathbb{W}\Rop{\MM_{aq} \mat{G}_q^H \mat{G}_q \MM_{aq}^H} \frac{\partial\vec{g}_a}{\partial \vec{{\edits{\gamma}}}_a} ~~+ \right. \\ & \qquad \left. \sum_{p \neq a} \bigg(\frac{\partial\vec{g}^{\edits{\dagger}}_a}{\partial \vec{{\edits{\gamma}}}_a}\bigg)^H
 \mathbb{W}\Lop{\MM_{pa}^H \mat{G}_p^H \mat{G}_p \MM_{pa}}
\frac{\partial\vec{g}^{\edits{\dagger}}_a}{\partial \vec{{\edits{\gamma}}}_a} \right]
\end{split}~~,
\end{equation}
and
\begin{equation}
\label{eq:slopes:overalljhr}
\JJ^H\RR =
\begin{bmatrix}
\displaystyle
\sum_{q \neq a} 2\operatorname{\mathbb{R}e}\bigg(\bigg(\frac{\partial\vec{g}_a}{\partial \vec{{\edits{\gamma}}}_a}\bigg)^H \vect\big(\mat{R}_{aq} \mat{G}_q \MM^H_{aq}\big)\bigg)
\end{bmatrix}.
\end{equation}

There is a slight departure from the solvers in the preceding sections. This is due to the fact that each derivative term is a $4 \times N_\mathrm{PPA}$ matrix. Consequently, their conjugate transposes are $ N_\mathrm{PPA} \times 4$ and $\tilde{\mat{H}}$ is a block diagonal matrix with $N_\mathrm{PPA} \times N_\mathrm{PPA}$ entries. For similar reasons, the entries of $\JJ^H\RR$ are $N_\mathrm{PPA} \times 1$ vectors.      

These expressions are sufficient for computing an update with respect to the phase slopes, but are, relatively speaking, still complicated. To further simplify them, we first need to examine the structure of the derivative terms:
\begin{equation}
\label{eq:slopes:derivs}
\begin{split}
\bigg(\frac{\partial\vec{g}_a}{\partial \vec{{\edits{\gamma}}}_a}\bigg)^H = 
\begin{bmatrix}
\bigg(\frac{\partial\vec{g}_a}{\partial \vec{a}_a}\bigg)^H \\
\bigg(\frac{\partial\vec{g}_a}{\partial \vec{b}_a}\bigg)^H \\
\bigg(\frac{\partial\vec{g}_a}{\partial \vec{c}_a}\bigg)^H 
\end{bmatrix} &=
\begin{bmatrix}
-i\nu\overline{g}_a^{XX}&0&0&0 \\
0&0&0&-i\nu\overline{g}_a^{YY} \\
-it\overline{g}_a^{XX}&0&0&0 \\
0&0&0&-it\overline{g}_a^{YY} \\
-i\overline{g}_a^{XX}&0&0&0 \\
0&0&0&-i\overline{g}_a^{YY} 
\end{bmatrix} \\
\bigg(\frac{\partial\vec{g}_a^{\edits{\dagger}}}{\partial \vec{{\edits{\gamma}}}_a}\bigg)^H =
\begin{bmatrix}
\bigg(\frac{\partial\vec{g}^{\edits{\dagger}}_a}{\partial \vec{a}_a}\bigg)^H \\
\bigg(\frac{\partial\vec{g}^{\edits{\dagger}}_a}{\partial \vec{b}_a}\bigg)^H \\
\bigg(\frac{\partial\vec{g}^{\edits{\dagger}}_a}{\partial \vec{c}_a}\bigg)^H 
\end{bmatrix} &= 
\begin{bmatrix}
\phantom{-}i\nu g^{XX}_a&0&0&0 \\
0&0&0&\phantom{-}i\nu g^{YY}_a \\
\phantom{-}itg^{XX}_a&0&0&0 \\
0&0&0&\phantom{-}itg^{YY}_a \\
\phantom{-}ig^{XX}_a&0&0&0 \\
0&0&0&\phantom{-}ig^{YY}_a
\end{bmatrix}
\end{split}.
\end{equation}

It should be immediately apparent that these derivative matrices contain three very similar blocks. They only differ by some multiplicative factor, which happens to be the coefficient of the parameter associated with that particular derivative. Additionally, the block associated with the intercept is identical to general phase-only case. This insight makes it possible to write out alternative expressions for the derivative terms:
\begin{equation}
\label{eq:slopes:kronderivs}
\begin{split}
\bigg(\frac{\partial\vec{g}_a}{\partial \vec{{\edits{\gamma}}}_a}\bigg)^H &= 
\vec{f} \otimes \bigg(\frac{\partial\vec{g}_a}{\partial \vec{\phi}_a}\bigg)^H\\
\bigg(\frac{\partial\vec{g}_a^{\edits{\dagger}}}{\partial \vec{{\edits{\gamma}}}_a}\bigg)^H &= 
\vec{f} \otimes \bigg(\frac{\partial\vec{g}^{\edits{\dagger}}_a}{\partial \vec{\phi}_a}\bigg)^H
\end{split}~~,
\end{equation}
where $\vec{f}$ is a vector of coefficients given by:
\begin{equation}
\vec{f} = 
\begin{bmatrix}
\nu \\
t \\
1 
\end{bmatrix}.
\end{equation}

We are now in a position to substitute these new expressions back into $\tilde{\mat{H}}$ and $\JJ^H\RR$ to obtain:
\begin{equation}
\begin{split}
\tilde{\mat{H}}  =
& \left[\sum_{q \neq a} \bigg(\vec{f} \otimes \bigg(\frac{\partial\vec{g}_a}{\partial \vec{\phi}_a}\bigg)^H\bigg)\mathbb{W}\Rop{\MM_{aq} \mat{G}_q^H \mat{G}_q \MM_{aq}^H} \bigg(\vec{f}^{\phantom{.}T} \otimes \bigg(\frac{\partial\vec{g}_a}{\partial \vec{\phi}_a}\bigg)\bigg) ~ + \right. \\ & \left. \quad \sum_{p \neq a} \bigg(\vec{f} \otimes \bigg(\frac{\partial\vec{g}^{\edits{\dagger}}_a}{\partial \vec{\phi}_a}\bigg)^H\bigg)
 \mathbb{W}\Lop{\MM_{pa}^H \mat{G}_p^H \mat{G}_p \MM_{pa}}
\bigg(\vec{f}^{\phantom{.}T} \otimes \bigg(\frac{\partial\vec{g}^{\edits{\dagger}}_a}{\partial \vec{\phi}_a}\bigg)\bigg) \right]
\end{split}
\end{equation}
and
\begin{equation}
\JJ^H\RR =
\begin{bmatrix}
\displaystyle\sum_{q \neq a} 2\operatorname{\mathbb{R}e}\bigg(\bigg(\vec{f} \otimes \bigg(\frac{\partial\vec{g}_a}{\partial \vec{\phi}_a}\bigg)^H\bigg) \vect(\mat{R}_{aq} \mat{G}_q \MM^H_{aq})\bigg)
\end{bmatrix}.
\end{equation}

The value of this substitution is not immediately apparent; first we need to invoke the mixed-product property of Kronecker products. For matrices $\mat{A}$, $\mat{B}$, $\mat{C}$ and $\mat{D}$ such that is possible to form the matrix products $\mat{A}\mat{C}$ and $\mat{B}\mat{D}$, the mixed-product property states:
\begin{equation}
\displaystyle (\mathbf {A} \otimes \mathbf {B} )(\mathbf {C} \otimes \mathbf {D} )=(\mathbf {AC} )\otimes (\mathbf {BD} ).
\end{equation}

When using this property, it is important to note that any matrix $\mat{A}$ can be written as either $\mat{A} \otimes \mat{1}$ or $\mat{1} \otimes \mat{A}$ where $\mat{1}$ is a $1\times1$ matrix containing a single $1$. We use this trick to simplify $\JJ^H\RR$:
\begin{equation}
\begin{split}
\JJ^H\RR &=
\begin{bmatrix}
\displaystyle\sum_{q \neq a} 2\operatorname{\mathbb{R}e}\bigg(\bigg(\vec{f} \otimes \bigg(\frac{\partial\vec{g}_a}{\partial \vec{\phi}_a}\bigg)^H\bigg) \bigg(\mat{1} \otimes \vect(\mat{R}_{aq} \mat{G}_q \MM^H_{aq})\bigg)\bigg) 
\end{bmatrix} \\
&=
\begin{bmatrix}
\displaystyle\sum_{q \neq a} 2\operatorname{\mathbb{R}e}\bigg(\vec{f}  \otimes \bigg(\bigg(\frac{\partial\vec{g}_a}{\partial \vec{\phi}_a}\bigg)^H \vect(\mat{R}_{aq} \mat{G}_q \MM^H_{aq})\bigg)\bigg)
\end{bmatrix}.
\end{split}
\end{equation}

Inspection of this expression reveals that is it the same as the generic phase-only case with the addition of the Kronecker product. This doesn't prevent us from using the same tricks we used in Section \ref{ssec:po} to write down a final simplified expression:
\begin{equation}
\JJ^H\RR =
\begin{bmatrix}
\displaystyle
\sum_{q \neq a} 2\operatorname{\mathbb{I}m}\bigg(\vec{f}  \otimes \mathrm{diag}(\mat{G}_a^H \odot \mat{R}_{aq} \mat{G}_q \MM^H_{aq}) \bigg)
\end{bmatrix}.
\end{equation}

This is a powerful result; any parameterised phase-only gain which is linear with respect to its parameters and has coefficients which do not vary with correlation will produce a $\JJ^H\RR$ of this form. Applying the mixed-product property to $\tilde{\mat{H}}$ we obtain:
\begin{equation}
\begin{split}
\tilde{\mat{H}}  =
& \left[\sum_{q \neq a} \vec{f}\vec{f}^{\phantom{.}T} \otimes \bigg(\bigg(\frac{\partial\vec{g}_a}{\partial \vec{\phi}_a}\bigg)^H \mathbb{W} \Rop{\MM_{aq} \mat{G}_q^H \mat{G}_q \MM_{aq}^H} \bigg(\frac{\partial\vec{g}_a}{\partial \vec{\phi}_a}\bigg)\bigg) ~~+ \right. \\ & \qquad \left. 
\displaystyle \sum_{p \neq a} \vec{f}\vec{f}^{\phantom{.}T} \otimes \bigg(\bigg(\frac{\partial\vec{g}^{\edits{\dagger}}_a}{\partial \vec{\phi}_a}\bigg)^H \mathbb{W} \Lop{\MM_{pa}^H \mat{G}_p^H \mat{G}_p \MM_{pa}}
\bigg(\frac{\partial\vec{g}^{\edits{\dagger}}_a}{\partial \vec{\phi}_a}\bigg)\bigg) \right]
\end{split}~~,
\end{equation}

Once again we are left with an expression which is familiar. To go further, we exploit the fact that the gains are diagonal and phase-only. As result, the approximations of Section \ref{ssec:po} apply and we can write:
\begin{equation}
\tilde{\mat{H}}  =
\begin{bmatrix}
\displaystyle
2 \sum_{q \neq a} \vec{f}\vec{f}^{\phantom{.}T} \otimes \bigg( \mat{I} \odot (\MM_{aq} \MM^H_{aq}) \bigg)
\end{bmatrix}.
\end{equation}

We are finally in a position where we can write the per-antenna update rule for the phase-slope case. First, however, we need to note a peculiarity of our derivation; we have omitted solution intervals. In the phase-slope case they are not optional as a slope across a single time-frequency slot is meaningless. \edits{Fortunately, including solution intervals is trivial and the general phase-slope update rule is given by}:

\begin{equation}
\begin{split}
\partial\vec{{\edits{\gamma}}}_a = 
& \bigg(\sum_{q \neq a,s} \vec{f}_s\vec{f}_s^{\phantom{.}T} \otimes \big( \mat{I} \odot (\MM_{aqs} \MM^H_{aqs}) \big)\bigg)^{-1} \\
& \qquad \bigg(\sum_{q \neq a,s} \operatorname{\mathbb{I}m}\Big(\vec{f}_s  \otimes \mathrm{diag}(\mat{G}_a^H \odot \mat{R}_{aqs} \mat{G}_q \MM^H_{aqs}) \Big) \bigg)
\end{split}~~.
\end{equation}

The addition of weights and directions to this expression is straightforward. 

\subsection{Solving for pointing errors}

One of the many sources of error facing interferometers is the mechanical pointing errors of the antennas that make up the array. Each antenna has an associated beam and the pointing errors offset this beam pattern. This has an adverse effect on observations which do not account for these errors in the calibration step. 

In order to solve for pointing error, we return to a scalar (non-polarised) derivation. The reason for this is intuitive; the pointing error is a physical effect that should be the same for all polarisations. As such, we assume that it is possible to constrain the pointing error using only Stokes I.

Pointing error differs slightly from the previous derivations in that it is inherently direction dependent - the beam affects each source differently. However, even though it is a direction-dependent effect, the underlying pointing error is not; we will not compute a pointing error per direction. This will introduce a summation over directions into our expressions. 

The NLLS problem can be written as follows in this case:
\begin{equation}
\min_{\{\bmath{\psi}_p\}}\sum_{pq}||r_{pq}||_F,~~
r_{pq} = d_{pq}-\sum_{d=1}^{N_D}v_{d,pq},
\end{equation}
where,
\begin{equation}
\label{eqn:beamv}
v_{d,pq} = e_{d,p} (\vec{\psi}_p) m_{d,pq} \overline{e}_{d,q}(\vec{\psi}_q).
\end{equation}

\edits{In equation \ref{eqn:beamv}, we have replaced our conventional gain ($g$) with $e_{d,p}$, which represents the effect of antenna $p$'s beam in direction $d$. These beam effects are parameterised by their associated pointing errors, $\vec{\psi}_p=[\Delta l_p, \Delta m_p]$.}

The parameter vector contains \edits{all of} the per-antenna pointing errors:
\begin{equation}
\vec{\Psi} = 
\begin{bmatrix}
\vec{\psi}_1, \dots, \vec{\psi}_{N_{\mathrm{A}}} 
\end{bmatrix}^T
= 
\begin{bmatrix}
\Delta l_1, \Delta m_1, \dots, \Delta l_{N_\mathrm{A}}, \Delta m_{N_\mathrm{A}}
\end{bmatrix}^T.
\end{equation}

In order to make use of our existing expressions for arbitrary parameterisations, we need to adapt equations \ref{eq:param:totaljhj} and \ref{eq:param:overalljhrsimplified} to work in the scalar case with the inclusion of directions. Fortunately, this is relatively straightforward and we obtain:
\begin{equation}
\label{eq:pe:overalljhr}
\JJ^H\RR =
\begin{bmatrix}
\displaystyle 
\sum_{q \neq a} \sum_{d = 1}^{N_\mathrm{D}} 2\operatorname{\mathbb{R}e}\bigg(\bigg(\frac{\partial e_{d,a}}{\partial \vec{\psi}_a}\bigg)^H  e_{d,q} \overline{m}_{d,aq} r_{aq}\bigg)
\end{bmatrix},
\end{equation}
and
\begin{equation}
\label{eq:pe:overalljhj}
\begin{split}
\tilde{\mat{H}}  =
& \left[ \sum_{q \neq a} \sum_{d = 1}^{N_\mathrm{D}}
|m_{d,aq}|^2 |\overline{e}_{d,q}|^2 
\bigg(\frac{\partial e^{\phantom{H}}_{d,a}}{\partial \vec{\psi}_a}\bigg)^H
\bigg(\frac{\partial e^{\phantom{H}}_{d,a}}{\partial \vec{\psi}_a}\bigg) ~~+ \right. \\ & \qquad \left. 
\displaystyle \sum_{p \neq a} \sum_{d = 1}^{N_\mathrm{D}}
|\overline{m}_{d,pa}|^2 |\overline{e}_{d,p}|^2
\bigg(\frac{\partial \overline{e}_{d,a}}{\partial \vec{\psi}_a}\bigg)^H
\bigg(\frac{\partial \overline{e}_{d,a}}{\partial \vec{\psi}_a}\bigg) \right]
\end{split}~~.
\end{equation}

The scalar-by-vector derivatives which appear in equations \ref{eq:pe:overalljhr} and \ref{eq:pe:overalljhj} are given by:
\begin{equation}
\label{eq:pe:derivs}
\begin{split}
\bigg(\frac{\partial e_{d,a}}{\partial \vec{\psi}_a}\bigg) &= 
\begin{bmatrix}
\displaystyle\frac{\partial e_{d,a}}{\partial \Delta l_a} & 
\displaystyle\frac{\partial e_{d,a}}{\partial \Delta m_a}
\end{bmatrix} \\
\bigg(\frac{\partial \overline{e}_{d,a}}{\partial \vec{\psi}_a}\bigg) &=
\begin{bmatrix}
\displaystyle
\frac{\partial \overline{e}_{d,a}}{\partial \Delta l_a} & 
\displaystyle
\frac{\partial \overline{e}_{d,a}}{\partial \Delta m_a}
\end{bmatrix}
\end{split}.
\end{equation}

These are $1 \times 2$ vectors, or $2 \times 1$ vectors under the conjugate transpose. The result is that the entries of $\tilde{\mat{H}}$ are $2 \times 2$ matrices. We can also make the observation that once again the two summations which make up $\tilde{\mat{H}}$ are conjugates of each other. As a result, we can write the following simplified form:
\begin{equation}
\label{eq:pe:overalljhjsimplified}
\tilde{\mat{H}}  =
\begin{bmatrix}
\displaystyle \sum_{q \neq a} \sum_{d = 1}^{N_\mathrm{D}}
2\operatorname{\mathbb{R}e}\bigg(|m_{d,aq}|^2 |\overline{e}_{d,q}|^2 
\bigg(\frac{\partial e^{\phantom{H}}_{d,a}}{\partial \vec{\psi}_a}\bigg)^H
\bigg(\frac{\partial e^{\phantom{H}}_{d,a}}{\partial \vec{\psi}_a}\bigg)\bigg)
\end{bmatrix}.
\end{equation}

In general, we do not have analytic expressions for the beam and cannot further simplify the derivative terms. In practice, these derivatives will usually be obtained numerically from beam cubes (holography or simulated).

Finally, the per-antenna update is given by:
\begin{equation}
\begin{split}
\partial\vec{\psi}_a = 
& \left( \sum_{q \neq a} \sum_{d = 1}^{N_\mathrm{D}}
\operatorname{\mathbb{R}e}\bigg(|m_{d,aq}|^2 |\overline{e}_{d,q}|^2 
\bigg(\frac{\partial e^{\phantom{H}}_{d,a}}{\partial \vec{\psi}_a}\bigg)^H
\bigg(\frac{\partial e^{\phantom{H}}_{d,a}}{\partial \vec{\psi}_a}\bigg)\bigg) \right)^{-1} \\ 
& \qquad \left( \sum_{q \neq a} \sum_{d = 1}^{N_\mathrm{D}} \operatorname{\mathbb{R}e}\bigg(\bigg(\frac{\partial e_{d,a}}{\partial \vec{\psi}_a}\bigg)^H  e_{d,q} \overline{m}_{d,aq} r_{aq}\bigg) \right)
\end{split}~~.
\end{equation}

In this case, as in the general direction-dependent case, we cannot avoid using the residuals. Note that the pointing error solver has not been implemented due to the difficulties in incorporating the beam and its derivatives. We present the mathematics to establish that such a solver is, however, possible.



\section{Implementing CubiCal}
\label{sec:impl}

In this section we will discuss the implementation of the derived calibration methods in a Python package named CubiCal. The code is open source and already freely available on GitHub (\url{https://github.com/ratt-ru/CubiCal}). Its documentation can be found at \url{http://cubical.readthedocs.io}.

\subsection{Language and optimisation}
CubiCal is implemented in a mixture of Python (using NumPy) and Cython. Python is widely used in the astronomy community due to its flexibility as a dynamic programming language and the rich support for scientific computing offered by libraries such as NumPy, SciPy and AstroPy. Unfortunately, Python code is interpreted at run time instead of statically compiled and this comes at the cost of performance. In particular, the Global Interpreter Lock (GIL) present in Python's standard CPython implementation restricts interpretation of pure Python code to a single thread. This severely limits Python's ability to exploit multi-core CPU architectures.

Several strategies can be employed to bypass this limitation. The sections of our code that are executed most frequently have been written in Cython, a Python-like language from which performant C code can be generated. The resulting C-extensions are not subject to the GIL, allowing multiple threads and processes to execute code simultaneously (across several CPU cores).

Whilst the use of Cython greatly accelerates our implementation, we go a step further by adopting a multi-processing strategy. Each process runs a separate CPython interpreter (with its own GIL) to solve independent chunks of the problem. This allows us to execute more pure Python code than a multi-threaded strategy, which is inherently limited to a single CPython interpreter and GIL.

Multi-processing can incur large overheads due to inter-process communication and concurrent memory access. We mitigate this problem by making use of shared memory. Data stored in shared memory can be accessed by multiple processes simultaneously, without extraneous memory copies. The conventional problems associated with shared memory, specifically multiple processes attempting to change the contents of a single memory address, are absent in our case due to the independence of the problems assigned to each process.

\subsection{Modularity}

CubiCal makes use of object-oriented programming (OOP) to construct a highly modular interface for the calibration routines. In practice, there is only one GN/LM solver for all the different types of solvable gains.

This is possible because each type of gain calibration has an associated Python object which we call a \textit{gain machine}. These gain machines contain all the relevant code for updating the gain which they describe.

Using gain machines makes the addition of additional types of gain calibration simple. Provided a gain machine object conforms to a set of predefined abstract requirements it can be called by the solver routine to update a gain. 

\subsection{Data handling}
\label{ssec:datahandling}

CubiCal's primary data source is the measurement set. The interface between CubiCal and the measurement set is handled by the python-casacore package (\url{http://github.com/casacore/python-casacore}). Whilst this is not particularly interesting on its own, it leads us into the next, almost mandatory requirement of any package which aims to deal with the huge volumes of data produced by new instruments.

To this end, we have implemented a chunking strategy which traverses the measurement set in blocks of time and frequency. This allows us to control the memory footprint of the code and is particularly useful on hardware with limited RAM. 

In \edits{fact}, the chunking scheme which CubiCal employs is two-tiered; the first level of chunking defines the quantity of data read from disk before performing calibration. This is usually a number of measurement set rows in their entirety. The second level of chunking defines how much data is assigned to each calibration process. 

The first of these chunking strategies also gives us the opportunity to further accelerate the code when employing multiprocessing. We handle all the disk related I/O (reads/writes) in a single process which can be interleaved with the calibration processes. 

In the simplest case of two processes, the first process (I/O) will read a single chunk of data. Once this read is complete, the calibration process will immediately begin calibration. The I/O process will also immediately begin reading the next chunk of data. This allows us to effectively offset our I/O overhead.  

The second tier of the chunking strategy also has a large impact on performance. If we attempt to calibrate too much data at a time, we will have very poor CPU cache performance. This is due to the fact that our arrays grow too large to be cached efficiently. Unfortunately this is often unavoidable as we are limited by the solution intervals which we employ - we cannot make our second tier chunks smaller than a single time/frequency solution interval.

CubiCal can also make use of a second data source - sky models from which it can predict model visibilities. This simulation is handled by the Montblanc package which includes an optimised (for both CPU and GPU) implementation of the RIME. This is also interleaved with calibration when using multiprocessing and goes a long way to mitigating the cost of model prediction which has, particularly in the direction-dependent case, been a major calibration bottleneck.  

\subsection{Data structures}

One often neglected topic of implementation is the data structure that is employed. By making a sensible choice it is possible to both accelerate computation and simplify the code.

CubiCal's data structure is based on an observation made by \cite{salvini2014}. Specifically, they note that their update rule can be written as the product between certain rows and columns of a number of correlation matrices. This is also true for the solvers derived here and is best demonstrated with an example.

Let us consider a simple three antenna case including polarisation. We define the following model and data correlation matrices (with zeros corresponding to the autocorrelations):
\begin{equation}
\mat{\hat{M}} = 
\begingroup
\begin{bmatrix}
    \mat{0} & \mat{M}_{12} & \mat{M}_{13} \\
    \mat{M}_{21} & \mat{0} & \mat{M}_{23}\\
    \mat{M}_{31} & \mat{M}_{32} & \mat{0}
\end{bmatrix} =
\begin{bmatrix}
    \mat{0} & \mat{M}_{12} & \mat{M}_{13} \\
    \mat{M}^H_{12} & \mat{0} & \mat{M}_{23}\\
    \mat{M}^H_{13} & \mat{M}^H_{23} & \mat{0}
\end{bmatrix},
\endgroup
\end{equation}

\begin{equation}
\mat{\hat{D}} = 
\begingroup
\begin{bmatrix}
    \mat{0} & \mat{D}_{12} & \mat{D}_{13} \\
    \mat{D}_{21} & \mat{0} & \mat{D}_{23}\\
    \mat{D}_{31} & \mat{D}_{32} & \mat{0}
\end{bmatrix} =
\begin{bmatrix}
    \mat{0} & \mat{D}_{12} & \mat{D}_{13} \\
    \mat{D}^H_{12} & \mat{0} & \mat{D}_{23}\\
    \mat{D}^H_{13} & \mat{D}^H_{23} & \mat{0}
\end{bmatrix},
\endgroup
\end{equation}
where each entry is a $2\times2$ block. We also define a three antenna gain vector of matrices:
\begin{equation}
\vec{\mat{G}} = 
\begin{bmatrix}
    \mat{G}_1 \\
    \mat{G}_2 \\
    \mat{G}_3
\end{bmatrix}.
\end{equation}

We introduce a new operator, the \textit{aligned product}. For two vectors of matrices with $N$ matrix elements, $\vec{\mat{A}}$ and $\vec{\mat{B}}$, their aligned product is given by:
\begin{equation}
\vec{\mat{A}} \diamond \vec{\mat{B}} =  
\begingroup
\begin{bmatrix}
    \mat{A}_1 \mat{B}_1\\
    \vdots \\
    \mat{A}_{N} \mat{B}_{N}
\end{bmatrix}.
\endgroup
\end{equation}

$\vec{\mat{A}}$ and $\vec{\mat{B}}$ are not necessarily aligned in the input - the product itself aligns them. This can be considered an element-wise product where the elements are $2\times2$ matrices.  

Returning to the update rule for a single term complex chain given in equation \ref{eq:pcomparison}, we can see that the update rule is completely equivalent to:
\begin{equation}
\mat{G}_{a,k+1} = 
\bigg[\sum \hat{\mat{D}}_{(a,:)} \diamond \vec{\mat{G}}^H \diamond \mat{\hat{M}}_{(:,a)}\bigg]
\bigg[\sum \mat{\hat{M}}_{(a,:)} \diamond \vec{\mat{G}} \diamond \vec{\mat{G}}^H \diamond \mat{\hat{M}}_{(:,a)}\bigg]^{-1}.
\end{equation}

This shows that, assuming we use correlation matrices as our basic data structure, our update rule can easily be implemented as products between matrices of matrices and vectors of matrices. Additionally, the summations are as simple as collapsing a single axis of resulting arrays. This approach also makes the inclusion of time/frequency solution intervals and directions easy; they are simply added as leading dimensions on the correlation arrays.

This approach has been found to work well in practice, with the caveat that storing the conjugate terms leads to an increased memory footprint. Given that we are free to tune the memory usage using the chunking strategy mentioned in Section \ref{ssec:datahandling}, this is not too problematic. Additionally, conjugation is a surprisingly costly operation as it relies on a memory copy. We avoid multiple conjugations by using this data structure.

\section{Results}
\label{sec:results}

This section presents the results of applying CubiCal to both simulated and real data. Additionally, we include the findings of some rudimentary benchmarking to establish the competitiveness of the implementation.   

\begin{figure*}	\includegraphics[width=\linewidth]{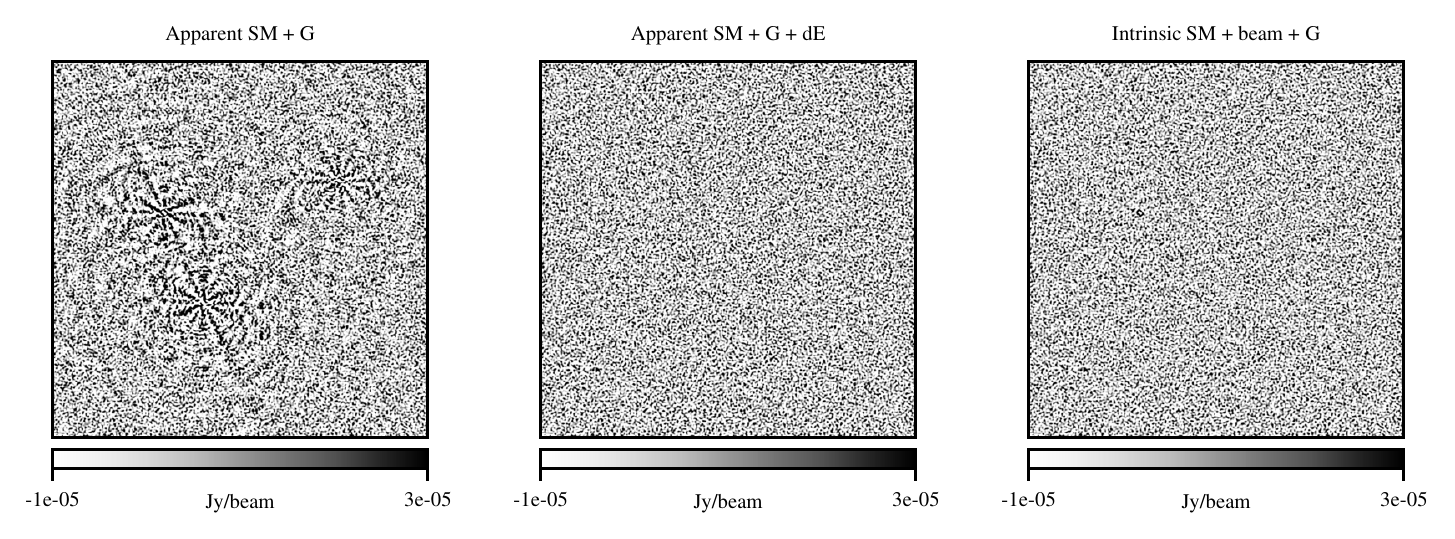}
    \caption{Residual maps of the beam-dominated patch after applying CubiCal to the simulated data using different gains and sky models (SMs). \edits{\textit{Left:} Residual map produced using an apparent sky model and a single direction-independent gain. \textit{Middle:} Residual map produced using an apparent sky model and both a direction-independent and direction-dependent gain. \textit{Right:} Residual map produced using an intrinsic sky model in conjunction with a beam model and a direction-independent gain.}}
    \label{fig:simpatch1}
\end{figure*}

\begin{figure*}
	\includegraphics[width=\linewidth]{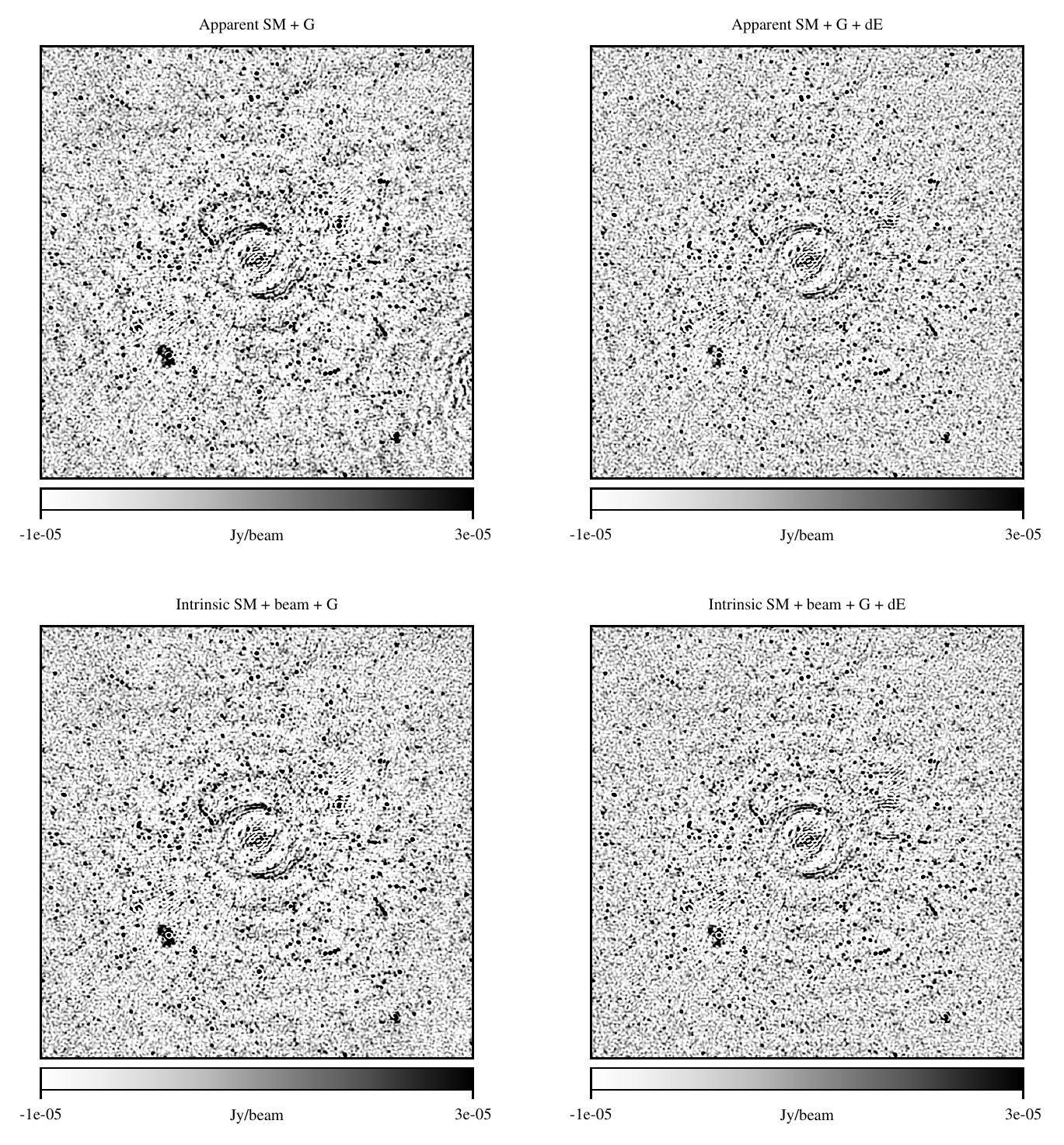}
    \caption{Residual maps of the field-centre patch after applying CubiCal to the observed data using different gains and sky models (SMs). \edits{\textit{Upper left:} Residual map produced using an apparent sky model and a single direction-independent gain. \textit{Upper right:} Residual map produced using an apparent sky model and both a direction-independent and direction-dependent gain. \textit{Lower left:} Residual map produced using an intrinsic sky model in conjunction with a beam model and a direction-independent gain. \textit{Lower right:} Residual map produced using an intrinsic sky model in conjunction with a beam model and both a direction-independent and direction-dependent gain.}}
    \label{fig:obspatch0}
\end{figure*}

\begin{figure*}
	\includegraphics[width=\linewidth]{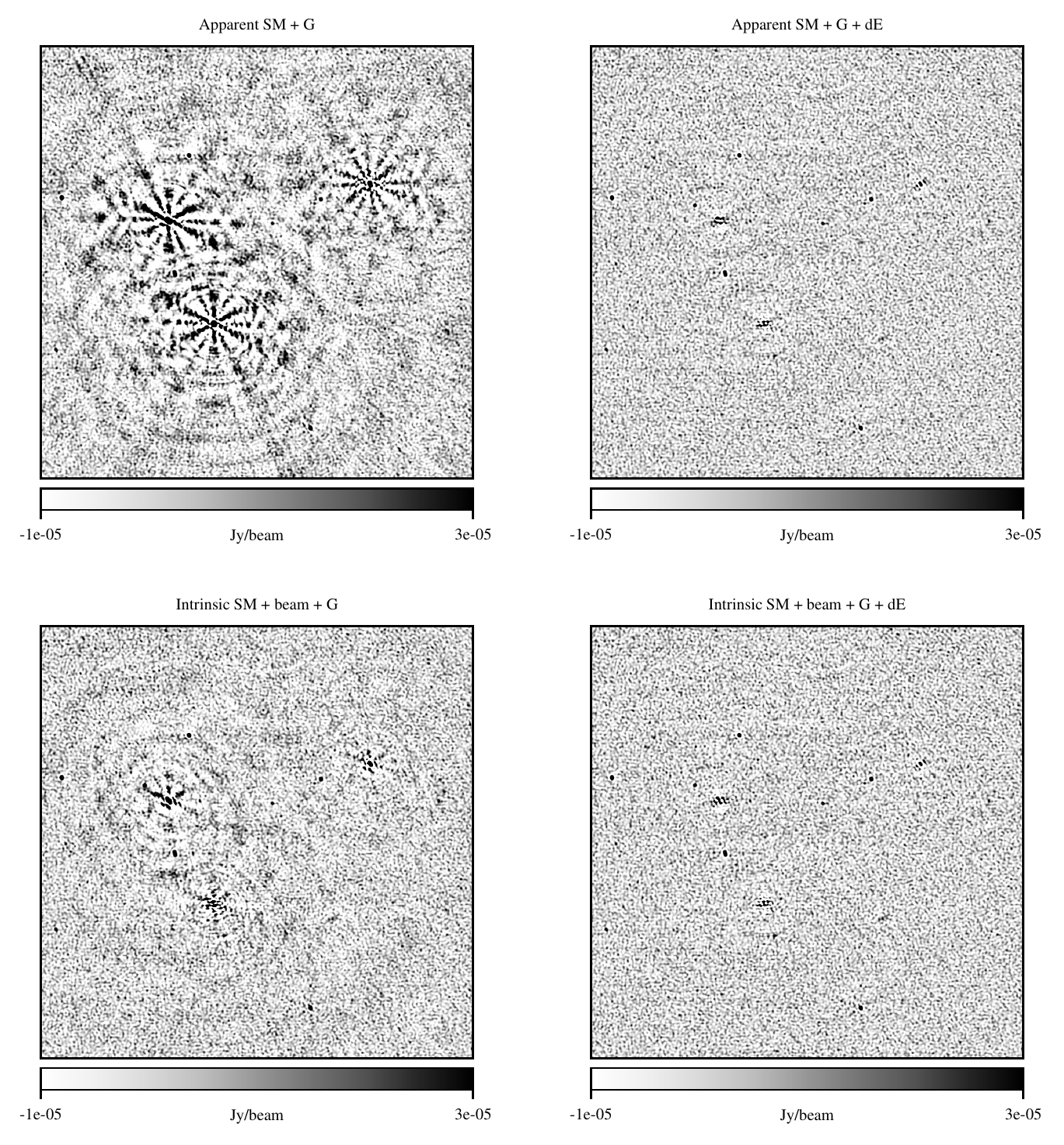}
    \caption{Residual maps of the beam-dominated patch after applying CubiCal to the observed data using different gains and sky models (SMs). \edits{\textit{Upper left:} Residual map produced using an apparent sky model and a single direction-independent gain. \textit{Upper right:} Residual map produced using an apparent sky model and both a direction-independent and direction-dependent gain. \textit{Lower left:} Residual map produced using an intrinsic sky model in conjunction with a beam model and a direction-independent gain. \textit{Lower right:} Residual map produced using an intrinsic sky model in conjunction with a beam model and both a direction-independent and direction-dependent gain.}}
    \label{fig:obspatch1}
\end{figure*}

\subsection{Application to simulated data}
\label{ssec:sims}

The target of our simulations is the field surrounding 3C147, a particularly bright compact source on which high dynamic range imaging is often performed. This is consistent with the real data we will be using, and allows us to make comparisons between the simulated and observed cases. In fact, we make use of both the same measurement set and sky model to ensure that our results are consistent. 

The measurement set in question is the product of a JVLA (Jansky Very Large Array) observation using the C-configuration. The target field was observed for a little under six hours with an integration time of five seconds. The bandwidth was divided into 64 channels with channel widths of 4MHz, ranging between 1.2665GHz and 1.5815GHz. An average of 26 antennas were present (unflagged) for the duration of the observation.

We made use of the MeqTrees software package \citep{noordam2010} to simulate visibilities corresponding to our sky model of the 3C147 field. The model in question contained 56 of the brighter sources in the field, obtained from prior self-calibration runs on real observations. These visibilities were corrupted by applying the JVLA primary beam (our sky model contains intrinsic flux), and including a varying complex gain. We also included noise, based on the SEFD (system equivalent flux density) of the JVLA at L-band.

Three different experiments were performed using the simulated data, and we compare the results by qualitative assessment of the residual images. We have selected two approximately 0.5 degree square patches of these residual images as being of particular interest. The first contains the field centre and the location at which 3C147 itself is subtracted. The second is away from the field centre in an area of the sky that is severely affected by the primary beam. This second patch contains three particularly troublesome sources which, in the absence of a beam, usually require direction-dependent gains. Note that in the simulated case, we omit the images of the first patch - they are entirely noise-like and of no particular interest.

The first experiment we performed was the most rudimentary. CubiCal was supplied with an apparent sky model (intrinsic fluxes modified by the primary beam) and we solved for a single, direction-independent gain term (G) on a (1,1) time-frequency solution interval. The patch containing the troublesome trio appears in the left-most panel of Fig.~\ref{fig:simpatch1}. Whilst they have been partially subtracted, there are still very noticeable artefacts in the residual map. This is consistent with the fact that those sources are subject to primary beam effects which cannot be adequately modelled with a single direction-independent gain. 

The second experiment is an extension of the first. We continue to use an apparent sky model, but use CubiCal to solve for both a direction-independent (G) term and a direction dependent (dE) term. The dE is solved on a (16,16) time-frequency solution interval to improve SNR and reduce over-fitting. The sky model contains ten sources flagged for dE solutions in seven unique directions. Using Montblanc, CubiCal simulates visibilities for each direction independently before performing calibration. The middle panel of Fig.~\ref{fig:simpatch1} contains the residual image after the application of direction-dependent gains. The artefacts which were obvious in the result of the first experiment have been eliminated.

The third and final experiment which we performed on the simulated data differs from the first two in that it makes use of the intrinsic sky model and a primary beam model obtained from EM simulations \citep{brisken2003}. As the beam is the direction-dependent effect which we included when corrupting the data, we expect that incorporating it in the calibration will make direction-dependent gains unnecessary. This is indeed the case, as is clear from the right-most panel of Fig.~\ref{fig:simpatch1}. The obvious artefacts have been eliminated and we can see that the second and third experiments produce very similar results. The advantage of incorporating the beam is that it can be applied to all the sources in the model without additional degrees of freedom. On the other hand, solving for a dE term on every source is degenerate and, in most cases, computationally intractable. However, as it is not always possible to accurately model the primary beam, it is promising that we can achieve similar results by applying a dE term.   



\subsection{Application to Real Data}
\label{ssec:obs}

Given CubiCal's success on simulated data, we moved on to applying it to the observation. We repeated all the experiments of subsection \ref{ssec:sims} using the observed data. The results of the these experiments appear in Fig.~\ref{fig:obspatch0} (the patch at the field centre) and Fig.~\ref{fig:obspatch1} (the beam-dominated patch).

The upper left panels of the figures correspond to the case where we use an apparent sky model and a single direction-independent gain term. Unlike the simulated case, the residuals in the patch at the field centre are not noise-like. This is due to the fact that the model is incomplete and may contain minor errors. 3C147 is particularly difficult to model accurately as it is slightly extended. This, coupled with its high brightness, leads to the slight over-subtraction visible in the image. The artefacts present in the beam-dominated patch are already familiar, and appear consistent with the simulated case. 

In the upper right panels of the figures, we see the results of applying a direction-dependent term. In the field centre patch, the improvement is subtle but manifests as a slight reduction in the overall noise. Artefacts introduced by sources outside the patch are noticeably reduced. The improvement in the beam-dominated patch is once again remarkable. Although there is slight over-subtraction at the positions of the troublesome trio, the artefacts are effectively eliminated. There are several reasons why the sources could \edits{be} over-subtracted. Once again, the model is neither perfect, nor complete. Additionally, the dE term is solved over a solution interval. This leads to a piecewise approximation of the underlying gain which cannot perfectly capture its behaviour. 

The results of the third experiment appear in the the lower left panels. In the case of the field centre, the application of the beam doesn't lead to a massive improvement. This is not surprising, as the effect of the primary beam is limited near the field centre. Close inspection of the image does, however, suggest a slight reduction in the overall noise as well a slight reduction in the number of artefacts. The improvement in the beam-dominated region is once again clear, though with an interesting difference. Unlike the simulated case, where the beam we introduced when corrupting the data is identical to the beam we apply during the model visibility prediction, there are still visible artefacts. This is due to the fact that the beam model we use, which is obtained from EM simulations \citep{brisken2003}, is not perfect. Consequently, it cannot entirely remove the effects of the primary beam. Additionally, unlike the simulated case, there could be other direction-dependent effects which affect the sources (atmospheric effects, for example). Thus, it is reasonable to apply a dE term even when using the beam.

To this end, we perform a fourth experiment. Like the third, we make use of an intrinsic sky model and apply the beam during the model visibility prediction step. We then solve for a direction-independent G term and direction-dependent dE term in the same way that we did in the second experiment. The fourth and final panel in the figures corresponds to this case. At the field centre, the only obvious sign of improvement is the superior subtraction of the bright source a little above and to the right of 3C147. This source was subject to the application of a direction-dependent gain, so this improvement makes sense. In the beam-dominated patch, the remaining artefacts are successfully removed and the final image is very similar to that obtained without the beam.

\subsection{Performance}
\label{ssec:perf}

\begin{figure}
	\includegraphics{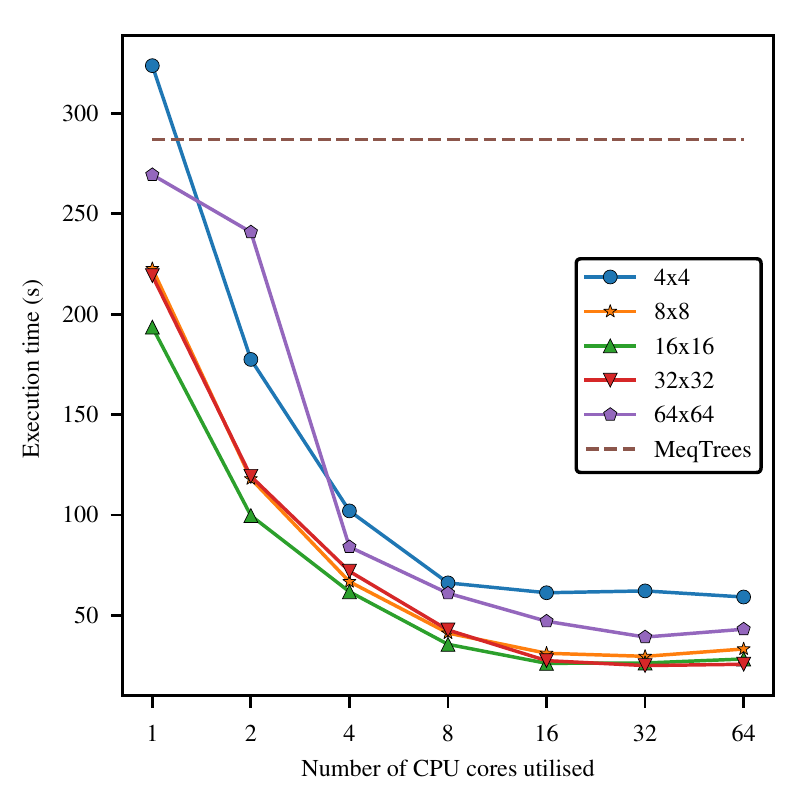}
    \caption{\edits{Execution speed as a function of number of processes (CPU cores) for different time by frequency chunk sizes.}}
    \label{fig:benchmarks}
\end{figure}

\begin{figure}
	\includegraphics{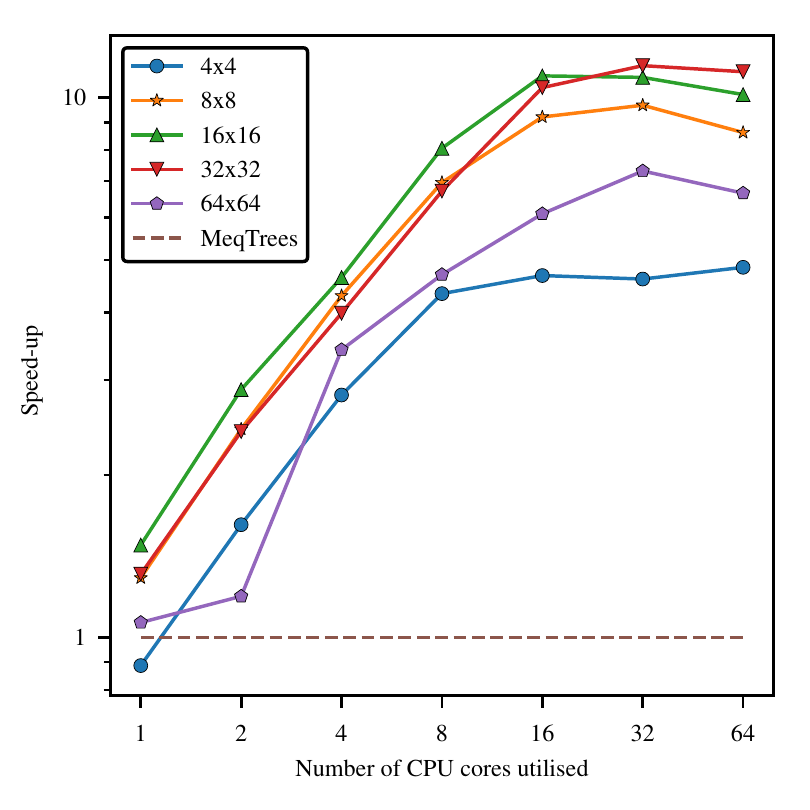}
    \caption{\edits{Speed-up relative to MeqTrees reference implementation as a function of number of processes (CPU cores) for different time by frequency chunk sizes.}}
    \label{fig:benchmarks2}
\end{figure}

Having established that CubiCal works and is successful in calibrating real data, we will present some rudimentary benchmarking results. 

It is, unfortunately, quite challenging to set up a fair comparison between several of the calibration tools currently available. Few of them implement the same algorithm, and even those that do may not have the same input parameters or parallelism. Ultimately, we settled on using MeqTrees as our reference point, as its full-polarisation StefCal implementation is fundamentally the same as a basic direction-independent G term in CubiCal.

Calibration in MeqTrees does not support any form of parallelism outside of its model prediction step, so we cannot make any claims about its performance scaling. It does, however, give us a good estimate of what would be considered typical speed for an existing tool. 

In order to gain some insight into the CubiCal's performance scaling, we need to return to the ideas of subsection \ref{ssec:datahandling}. CubiCal uses multiprocessing for parallelism, with each processes being assigned a block of data with some time and frequency dimension. These dimensions, or chunk size, have a large effect on CubiCal's performance. 

We once again made use of the JVLA observation of the 3C147 field. The observed and model visibility columns of the measurement set each contain around $3.5 \times 10^8$ visibilities, and each occupies 2.8GB on disk. Whilst this is not a large problem, it is sufficient to demonstrate CubiCal's performance. Note that in this instance, we calibrate using pre-simulated data stored in the measurement set. This is ensures that we do not profile Montblanc.

We calibrated the data several times, varying both the number of CPU cores (equivalent to the number of processes) and size of the time-frequency chunks, and compared the resulting execution times (wall times). Our test machine was running two Intel(R) Xeon(R) E5-2695 v4 CPUs, and 512GB of 2400MHz memory. The results of this experiment appear in Fig.~\ref{fig:benchmarks} \edits{and Fig.~\ref{fig:benchmarks2}}.

Each line corresponds to a different chunk size, as indicated in the legend. The dotted line near the top of \edits{Fig.~\ref{fig:benchmarks}} was the time taken for a MeqTrees StefCal calibration using identical input parameters. It is clear that the multiprocessing works and that CubiCal's performance does increase as we allocate it more resources. \edits{This is even clearer in Fig.~\ref{fig:benchmarks2}, which shows the speed-up obtained in CubiCal relative to the MeqTrees implementation.} One subtlety which is omitted from both figures is that, in all but the single CPU case, we used one additional process for I/O. This was done to ensure that we were doubling the available compute at each data point. 

The figures also shows that we have the best performance for time-frequency chunks with dimensions (16,16) and (32,32). This is due to the fact that at those chunk sizes, each process is doing sufficient work to minimise overhead, and we have good cache behaviour. This is clearly not the case for the (64,64) chunks. There is a sudden degradation in performance and this can be attributed to poorer cache behaviour - the arrays which CubiCal uses internally no longer fit in the CPU cache and there is an increase in cache-misses. 

It is also interesting to note the asymptotic behaviour above 16 processes. In the case of the fastest chunk sizes, this asymptote lines up quite well with the time taken to read the measurement set columns; we cannot calibrate faster than we read the data from disk. However, it is promising that we reach this limit. This experiment was performed using conventional platter drives, but we do aim to test the code on SSDs (solid state drives) in the future.


\section{Conclusions}
\label{sec:conc}

A reformulation of radio interferometric gain calibration using complex optimisation has paved the way for the development of several new gain solvers and strategies. We have extended the existing formulation to chains of Jones terms, providing a tractable means of calibrating for several different gains without manipulating the data. 

We made use of the properties of parameterised gains to derive several specialised and accelerated solvers. In particular, we have developed solvers for true phase-only calibration, delay/rate, and pointing errors. Additionally, we have presented fairly generic expressions for deriving solvers for arbitrarily parameterised gains. Each of the derived solvers makes use of diagonal approximations to reduce the computational cost of performing the NLLS updates.

Several of the presented solvers have been implemented in a Python package called CubiCal. CubiCal has been optimised using Cython and multiprocessing, and has been designed to allow for the incorporation of additional solvers. Its underlying data structures are particularly conducive to the calibration problem and CubiCal has been shown to be substantially faster than its most comparable competitor.

We have successfully applied CubiCal, using a variety of different gains, to both simulated and real visibility data. This has also been verified by users who have obtained excellent results. The code is already freely available, and we hope that it will be adopted as the de facto tool for self-calibration as data volumes continue to grow.

Whilst we do intend to add additional solvers to CubiCal, there are also several other ways in which it could be extended. There are limitations associated with the $2 \times 2$ formalism we use throughout the paper which are not present in the $4 \times 4$  formalism. We could incorporate this into a future version of CubiCal, allowing us to include arbitrary parametrisations in our chains of Jones terms. We could also give further thought to parametrisations reliant on numerical differentiation, such as the beams in the pointing error case. 



\section*{Acknowledgements}

Jonathan Kenyon acknowledges the financial assistance of the South African SKA Project (SKA SA) towards this research (\url{www.ska.ac.za}). Oleg Smirnov's research is supported by the South African Research Chairs Initiative of the Department of Science and Technology and National Research Foundation. We would like to thank Dr Ian Heywood for his helpful bug reports and Dr Cyril Tasse for some helpful conversations when the project was in its infancy. The anonymous reviewer has our gratitude for their helpful comments and suggestions.




\bibliographystyle{mnras}
\bibliography{cubical} 








\bsp	
\label{lastpage}
\end{document}